\documentclass[%
reprint,
amsmath,amssymb,
aps,
prx,
]{revtex4-2}

\usepackage{pstricks}
\usepackage{graphicx}
\usepackage{dcolumn}
\usepackage{bm}
\usepackage{color}

\usepackage{placeins}

\newcommand{\defaultfigurewidth}{0.47\textwidth}
\usepackage{braket}

\begin{document}
	
	\preprint{APS/123-QED}

	\title{High-fidelity robust qubit control by phase-modulated pulses}

	\author{Marko Kuzmanovi\'c}
	\affiliation{InstituteQ and QTF  Centre  of  Excellence,  Department  of  Applied  Physics, School  of  Science,  Aalto  University,  FI-00076  Aalto,  Finland}
	\author{Isak Bj\"orkman}
	\affiliation{InstituteQ and QTF  Centre  of  Excellence,  Department  of  Applied  Physics, School  of  Science,  Aalto  University,  FI-00076  Aalto,  Finland}
	\author{John J. McCord}
	\affiliation{InstituteQ and QTF  Centre  of  Excellence,  Department  of  Applied  Physics, School  of  Science,  Aalto  University,  FI-00076  Aalto,  Finland}
	\author{Shruti Dogra}
	\affiliation{InstituteQ and QTF  Centre  of  Excellence,  Department  of  Applied  Physics, School  of  Science,  Aalto  University,  FI-00076  Aalto,  Finland}
	\author{Gheorghe Sorin Paraoanu}\thanks{sorin.paraoanu@aalto.fi}
	\affiliation{InstituteQ and QTF  Centre  of  Excellence,  Department  of  Applied  Physics, School  of  Science,  Aalto  University,  FI-00076  Aalto,  Finland}

	\date{\today}

	\begin{abstract}
		We present a set of \textit{robust} and \textit{high-fidelity} pulses that realize paradigmatic operations such as the transfer of the ground state population into the excited state and  arbitrary $X/Y$ rotations on the Bloch sphere. These pulses are based on the phase modulation of the control field. We implement these operations on a transmon qubit, demonstrating resilience against deviations in the drive amplitude of more than $\approx 20\%$ and/or detuning from the qubit transition frequency in the order of $10~\mathrm{MHz}$. The concept and modulation scheme is straightforward to implement and it is compatible with other quantum-technology experimental platforms. 
	\end{abstract}

	\maketitle

	\section{Introduction}
	
	Quantum control -- a toolbox of techniques enabling high-fidelity dynamical operations -- is an essential tool in modern quantum technologies. Perhaps the first example of quantum control is the 1932 Rosen-Zener $\mathrm{sech}$-function  design of the shape of the rate of rotation of the magnetic field in a double Stern-Gerlach experiment  \cite{RosenZener1932}.
	
	In contrast to designing the shape of the pulse envelope, changing its phase or frequency is a less explored avenue. While simple forms of modulation (linear, sinusoidal, square) have been studied experimentally \cite{Silveri_2017}, this has confined the use of this concept to repetitive passages, which have the benefit that interference effects can been observed in a straightforward way.
	However, recent advances in modern electronics have enabled the precise manipulation of the pulse phases in the time domain. This allows the formulation of quantum control schemes where the phase is an externally-controlled parameter.

	An outstanding problem that can be addressed by these methods is the realization of high-fidelity gates and other operations in superconducting qubits, one of the most promising platforms for quantum computing and simulation. 
	Compared to other well-established experimental platforms, superconducting qubits present an additional specific set of challenges. Since these are artificial atoms comprising several materials and complex geometries, it is in general not possible to provide a sufficiently accurate and complete mathematical model describing the system, leading to losses and unaccounted-for interactions.
	To combat this several concepts have been proposed and applied recently: for example error mitigation (extrapolation to zero-noise limit or probabilistic error cancellation) \cite{Gambetta_2017,Benjamin_2017, Gambetta_2019}, Pauli and Clifford twirling \cite{Emerson_2008,Emerson_2012}, derivative removal by adiabatic gate \cite{Motzoi_2009,Gambetta_2011,Motzoi_2013}, dynamical decoupling \cite{Oliver_2011, Lidar2018, Lidar_2022}, counterdiabatic methods \cite{Chen_2010,Torrontegui_2013,Vepsalainen_2018,Vepsalainen_2019,Dogra_2022}, composite pulses \cite{Torosov_2014, Ivanov_2015, Sugny_2020, Torosov_2020}, and more recently reinforcement learning \cite{Henson_2018, Bukov_2018, An_2019, Neven_2019, Brown_2021,SauvageMintert2022,Giannelli_2022,ding2020breaking,ai2021experimentally,SauvageMintert2022}.
	There is however no universal solution, as each of these methods comes with it's own disadvantages. For example, machine-learning techniques typically require discretized forms of multiple control parameters, which complicates their synthesis by standard control systems.
	While this problem may be alleviated in the future by the use of cryogenic control systems, for example Josephson arbitrary wave-form synthesizers \cite{Sirois_2020}, such pulses may still have a power spectrum leading to spurious excitations in a larger, frequency-crowded device. Composite pulses, while effective for mitigating errors in the Rabi frequency, are in general not protecting against frequency shifts and the concatenated series of pulses take a long time, allowing decoherence to take its toll. 
	
	Also, techniques that require a very precise shape optimization of pulses may not be so effective,  
	since microwave pulses will be inevitably distorted when transmitted to the qubit, and finding the exact transfer function requires further extensive calibrations \cite{Oliver_2013, Fedorov_2019, DiCarlo_2020}. 
	Another solution could be to use a closed-loop approach based on randomized benchmarking of a subset of gates \cite{Martinis_2014}, a method that can be extended to include leakage control with tens of parameters to optimize \cite{Filipp_2021}. However, as the shape of the pulse becomes more complex, a large number of parameters are needed, and the optimization time may increase significantly \cite{Filipp_2021}.

	Here we show that by chirping the frequency of the pulses according to relatively simple and smooth functions, we can achieve gates that are robust against both amplitude and frequency errors. We restrict the power of the microwave pulse in order to avoid exciting other modes due to frequency crowding, also to limit the effect of nonlinearities in microwave components (especially mixers at room temperature) and in the on-chip circuit elements (other nearby qubits, modes, etc) that may lead to frequency shifts that are not accounted for in calibration. Combined with the requirement of minimizing the time of the operation, this inevitably leads to pulses that are close to rectangular.
	
	We demonstrate this by implementing two paradigmatic operations: transfer of population from one level to another and arbitrary rotations on the Bloch sphere. In both cases, the 
	experimental data are supported by simple theoretical models without any free parameters and by numerical simulations.
		
	These control techniques  are validated in a setup comprising a transmon qubit, although the methods developed here are, in principle, hardware-agnostic. Besides robustness and high-fidelity, which will be substantiated further, a great advantage of our scheme is simplicity: we work with a small number of optimization parameters - the rate of the frequency variation for the first task and a few Fourier coefficients for the second one, which makes the optimization very efficient numerically. 
		
	Our results have immediate applications in a wide array of quantum-information tasks realized with superconducting circuits.
	Population transfer is an important operation in quantum information processing, as it apears in various contexts - in thermometry protocols \cite{Sultanov2021}, qubit calibration, and quantum engines such as batteries and single-atom lasing.
	For example, calibration of a large number of qubits in a superconducting quantum processor is a time-costly process, and it has to be done frequently due to uncontrolled frequency drifts of the qubits and of the control electronics. 
	Therefore, having resilience to small drifts in system and/or control parameters is highly beneficial.

	The single-qubit gates with phase-modulated pulses have straightforward usage in gate based quantum computing, where the robustness can mitigate fluctuations of control parameters, e.g. over long time scales.
	Another application could be in NMR or systems with global control \cite{shapira2023programmable}, where the same drive pulse acts on more than one site: the amplitude robustness would enusre a homogeneous drive even with variations in the drive-atom coupling strength.
	
	The paper is organized as follows: in Section \ref{methods} we introduce the key concepts underlining the experiments. The transfer is addressed in Section \ref{transfer}, while a general rotation is designed and implemented in Section \ref{rotation}. We recapitulate the main findings of the paper in Section \ref{discussion}. 
	
	\section{Methods} \label{methods}

	When subject to a drive field with a (time-dependent) frequency $\omega_d(t)$ and (Rabi) amplitude $\Omega(t)$, in the frame co-rotating with the drive, the Hamiltonian of a qubit becomes:
	
	\begin{equation}
		H = \frac{\hbar}{2}\begin{pmatrix}
			-\Delta(t) & \Omega (t) \\
			\Omega(t) &	\Delta(t)
		\end{pmatrix},
		\label{eq:hamiltionian}
	\end{equation}	
	where $\Delta(t)$ is the \textit{instantaneous} detuning between the qubit ($\omega_q$) and the drive frequency $\Delta(t) = \omega_d(t) - \omega_q + t \partial \omega_d(t)/\partial t$. Specifically, starting with the Hamiltonian $H = (\hbar \omega_q /2)\sigma_z + \hbar \Omega \cos (\omega_d t) \sigma_x$, we perform a rotation  $R = \exp (-i \omega_d(t) t \sigma_z /2)$ around the $z$ axis. The Hamiltonian transforms as $H \rightarrow R HR^{\dag} - i \hbar R(\partial R^{\dag}/ \partial t)$, and with the subsequent use of the rotating wave approximation one gets Eq.~(\ref{eq:hamiltionian}).	Here we take the Rabi coupling to be real, without loss of generality, as any phase modulation can be re-written in terms of a frequency modulation though the choice of the rotating frame. The control parameters of the drive are now the time-dependent Rabi frequency $\Omega (t)$ and the detuning $\Delta(t)$.
	
	The qubit used in our experiments is a transmon device with a first transition frequency of $f_{01}=7.27~\mathrm{GHz}$.  The charging  energy of the shunting capacitor for this sample was $E_C\approx 340~\mathrm{MHz}$ and the measured relaxation times were obtained as  $T_1\approx 7~\mathrm{\mu s}$, $T_2^{\rm Ramsey} \approx 5~\mathrm{\mu s}$. The sample was thermally anchored to the mixing chamber of a dilution refrigerator with 10 mK base temperature and connected to room-temperature microwave electronics using $\approx 70~\mathrm{dB}$ of attenuation on the control and $\approx 90~\mathrm{dB}$ of attenuation on the readout lines. 
			
	Our control scheme requires modulating the frequency of the pulses, which can be done in a straightforward way by mixing the local oscilator (LO) tone with a modulated intermediate-frequency (IF) signal generated by an arbitrary waveform generator (AWG). For high-coherence qubits, the fidelities and gate errors are then limited mostly by technical specifications that improve all the time, such as the sampling rate and the characteristics of the mixers.
	Here we used a typical IQ mixer setup employing a Marki IQ-4509 mixer, with the intermediate frequency signals  being generated by a Tektronix 5204 AWG (sampling rate $5~\mathrm{GS/s}$, bandwidth $\approx 2~\mathrm{GHz}$) or by a Quantum Machines OPX+ system (sampling rate $1~\mathrm{GS/s}$, bandwidth $\approx 350~\mathrm{MHz}$).  With this setup it is straightforward to generate a signal with a time-dependent amplitude $\Omega(t)$ and frequency $\omega_d(t)$. Starting with a local oscillator $\mathrm{LO}=e^{i \omega_{\mathrm{LO}} t}$ and setting $I + iQ \propto e^{i \omega_{\rm IF} t}$, one ends up with a signal $\mathrm{RF} = e^{i (\omega_{\mathrm{LO}} + \omega_{\mathrm{IF}}) t}$. Therefore, if the intermediate-frequency signal is $I + iQ = \Omega(t) e^{i (\omega_d(t) - \omega_{\mathrm{LO}}) t}$ a  signal $\mathrm{RF} = \Omega(t)e^{i \omega_{d}(t)t}$ will be generated with the desired envelope $ \Omega(t)$. In this way, any $\Delta(t)$ can be generated by an appropriate choice of $\omega_d(t)$. Even the (comparatively) small bandwidth of the latter generator did not affect the performance of the pulses greatly, hence the proposed control scheme should be possible to implement on most control hardware.	
	
	Readout is perfomed dispersively, using a higly-detuned superconducting resonator coupled to the qubit,	in the averaged regime. After post-processing of the acquired data we obtain a readout noise of $\sigma(P_1) \approx 0.3\%$.
	
	In order to fully characterize the quantum gates we perform  quantum process tomography (QPT) following the standard procedure \cite{nielsen_chuang_2010}: the qubit is initialized in the states $|0\rangle$, $|1\rangle$,  $(|0\rangle + |1\rangle)/\sqrt{2}$, and	$(|0\rangle + i|1\rangle)/\sqrt{2}$. Qubit, in each of these initial states is then allowed undergo same quantum gate operation and respective final states are obtained. Quantum state tomography of the respective final states is then performed in each case using three operations: identity ($\mathbb{I}$),  $(\pi/2)_y$ rotation and $(\pi/2)_x$ rotation giving rise to the expectation values of the Pauli operators $\sigma_z$, $\sigma_x$, and $\sigma_y$ respectively. These experimental results are then used to reconstruct the $\chi_{\rm exp}$  process matrix \cite{nielsen_chuang_2010}, subject to positivity constraints for both the experimental density matrices as well as the reconstructed process matrix.

	The fidelity of the quantum gate is measured by $F = {\rm Tr}(\chi . \chi_{\rm exp})$, which corresponds to the average gate fidelity \cite{nielsen2002simple}, where $\chi$ is the theoretically expected process matrix.
		
	Using this setup we implement two fundamental operations: population transfer and arbitrary rotations. First, we explore control schemes leading to a \textit{population transfer} of the qubit from the ground to the excited state. Further, we demonstrate how a similar modulation scheme can be used to implement \textit{amplitude-robust} arbitrary $X/Y$ rotations on the Bloch sphere.
		
	While optimizing the pulse parameters the cost function $\|U_\mathrm{realized} - U_\mathrm{target}\|_{\rm F}$ was minimized, where  $\| M \|_{\rm F} = \sqrt{\sum_{i,j=1}^{2}|m_{ij}|^2} = \sqrt{\mathrm{Tr}\left( M^{\dag}M\right)}$  is the Frobenius norm. While the full form is necessary for an arbitrary operation, for a population transfer it is sufficient to have the matrix element $U_{10}=\bra{1}U\ket{0}$ satisfy $|U_{10}|=1$, which is directly accessible by applying the pulse to a qubit initialized in the ground state and measuring the population of the excited one.

	\section{Robust population transfer}\label{transfer}

	The standard way of realizing population transfer is by a simple Rabi $\pi$ pulse; however, Rabi pulses are sensitive to errors both in amplitude and frequency. To find a robust pulse, we start our construction with the observation that ideal adiabatic processes are immune to variations in the path of the control parameters. Indeed, the instantaneous eigenenergies of the Hamiltonian Eq.~(\ref{eq:hamiltionian}) are $E_{\pm}= \pm \frac{\hbar}{2}\sqrt{\Omega^2 + \Delta^2}$, corrresponding to the eigenstates $\ket{E_{-}}= \cos(\Theta) |0\rangle  -\sin(\Theta) |1\rangle  $ and $\ket{E_{+}}= |\sin(\Theta)|0\rangle + \cos(\Theta) |1 \rangle $, where the mixing angle $\Theta$ is defined as  $\tan(\Theta) =\left(\sqrt{\Omega^2 + \Delta^2}+ \Delta\right)/\Omega$ and $|0\rangle = (0,1)^{\rm T}$ and $|1\rangle = (1,0)^{\rm T}$ are the ground and excited state. 
	If initially / finally $|\Delta| \gg \Omega$ the eigenstates of the Hamiltonian are close to the ground ($\ket{0}$) and excited ($\ket{1}$) states, and if $\Delta$ changes sign they transform from $\approx \ket{0}$ to $\approx \ket{1}$ (and vice versa). According to the adiabatic theorem if the rate of change of $\Delta$ is slow  the system will follow the instantaneous eigenstates, leading to a population transfer between $\ket{0}$ and $\ket{1}$.
	
	In practice however, the transfer time $T$ is inevitably finite - which strictly speaking breaks the adiabaticity - and, moreover, a specific choice of trajectory in the parameter space has to be made. The simplest such choice is a linear chirp of frequency, which requires only one control parameter (the speed of the chirp). This has experimental advantages (easiness of programming the waveform) as well as theoretical ones - since it implements the celebrated
	Landau-Zener-St\"uckelberg-Majorana (LZSM) model, which can be solved analytically \cite{Shevchenko_2023}. We note that early theoretical works \cite{Nori_2008} proposed to use ac-Stark shifts to modulate the phase appropriately. In the present experiment we realize this task by implementing the phase modulation directly, leveraging on the mixing  methods described in the previous section. In contrast to the present work, the observation of LZSM in circuit QED typically involves direct modulation of the qubit frequency using a strong rf field \cite{Oliver_2005, Mika_2006, Silveri_2015}. 
	
	We  parametrize the detuning as $\Delta(t) = t \dfrac{2\Delta_{\rm max}}{T}$, where $\Delta_{\rm max}$ is the modulation depth and $T$ is the pulse duration, while the drive frequency is modulated as $\omega_d(t) = \omega_q + t \dfrac{\Delta_{\rm max}}{T}$ and $t\in[\frac{-T}{2},\frac{T}{2}]$. Note that the factor of $2$ difference between the modulation of $\omega_d$ and $\Delta$ is a result of the definition of the instantaneous detuning.
	For such a parametrization it is natural to measure the Rabi frequency $\Omega$ and detuning $\Delta_{\rm max}$ in units of $\Omega_{2\pi} = \frac{2\pi}{T}$, the Rabi frequency needed for a $2\pi$ rotation without detuning, as the frequency scale is set by $T^{-1}$.

	We start the design of the phase-modulated pulse by studying first  a rectangular-shape pulse, see Appendix A. This already results in robustness with respect to the pulse amplitude. To achieve robustness also with respect to detuning, we show that it is sufficient to ``soften" the edges of the pulse by using a super-Gaussian shape. This shape reduces the nonadiabatic excitations and, simultaneously, allows the pulse to still be confined in the time-domain (thus avoiding the effect of decoherence). The concept is similar to the rapid adiabatic passage process \cite{Abragam1961, Vitanov2001, Shore2011}, where the shape of the pulse envelope is identified by conditions that achieve the suppression of nonadiabatic excitations. For example, a simple idea is to just eliminate the transition points by imposing that the gap between the instantaneous eigenstates remains constant \cite{Guerin_2002,Lacour_2008}. 
	
	Here we adopt a similar strategy, identifying the transitions at the beginning and end of the pulse, and suppressing them by a suitable choice smoothing the pulse edges that keeps the transition rate below a certain threshold. If $\Omega$ is time-dependent and vanishes at $t=\pm T/2$, at the beginning of the drive, the mixing angle is zero $\Theta (-T/2) = 0$ and the lower instantaneous eigenstate $|E_{-}\rangle$ of the Hamiltonian Eq.~(\ref{eq:hamiltionian}) coincides with the ground state $|0\rangle$ of the qubit. While at the end of the pulse ($t=T/2$), the mixing angle is $\Theta (T/2) = \pi/2$ and the lower eigenstate corresponds, up to a sign, to the excited state $|1\rangle$ of the qubit. The crossover happens at $t=0$, when the pulse is resonant with the qubit, with a gap of $\Omega(t=0)$ in the spectrum. For an appropriate choice of parameters, one can imagine that the qubit will adiabatically follow the lower eigenstate, resulting in a population transfer.

	To achieve this, we opted for the super-Gaussian envelope, given by $\Omega(t) = \Omega_0 e^{-\left(t /\tau\right)^4}$, with $\tau$ chosen such that $\Omega(t=\pm T/2)=0.01 \Omega_0$. 
	This choice meets the condition of an \textit{approximately} vanishing amplitude at $t=\pm T/2$, and has an under-the-curve area  of $\approx60\%$ compared to a rectangular pulse of the same duration, resulting in only a moderate increase of the peak amplitude $\Omega_0$ needed for the same rotation. Moreover, it has been shown  that Gaussian shapes are effective in reducing the leakage outside the computational space, even leading to less errors than some composite pulses \cite{Steffen_2003}.
	
\begin{figure}
		\centering
		\includegraphics[width=\defaultfigurewidth]{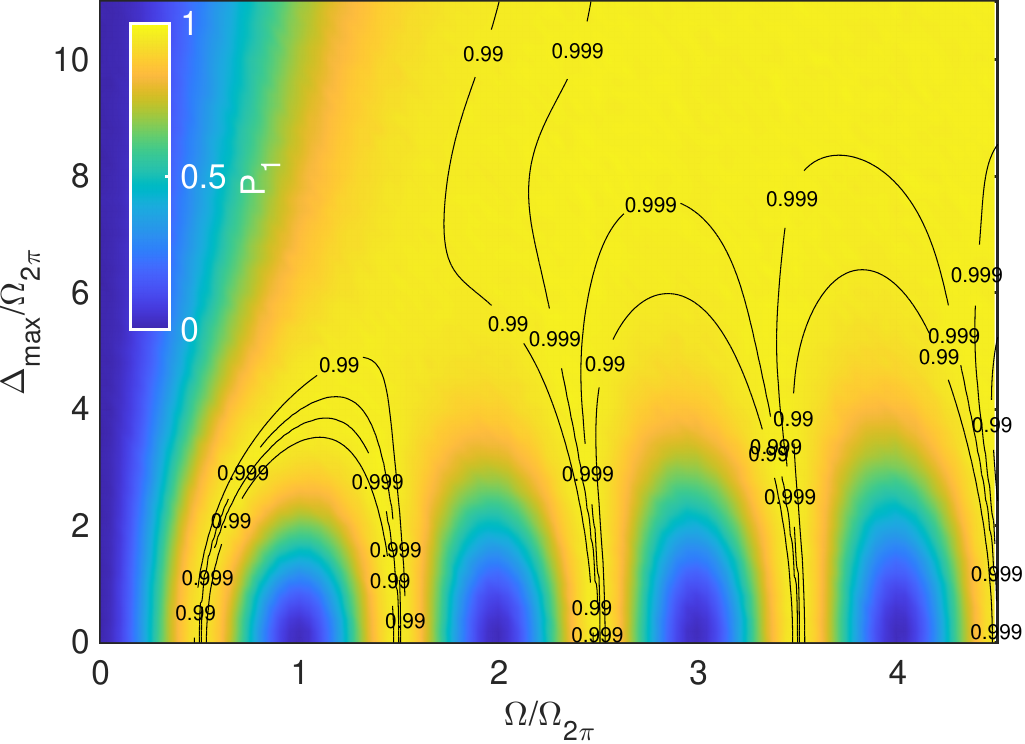}
		\caption{The experimentally observed population of $\ket{1}$, $P_1$, after applying the $(\Omega,\Delta_{\rm max})$ pulse with a super-Gaussian envelope (colored plot), along with the theoretical prediction (contour lines).			
		}
		\label{fig:supergaussian_envelope_amp_delta_max}
	\end{figure}

	The $(\Omega,\Delta_{\rm max})$ parameter space was explored numerically, as well as experimentally. 
	The pulse duration was $T=200~\mathrm{ns}$, corresponding to a $2\pi$ Rabi frequency of $\Omega_{2\pi} = 2\pi \times 5\mathrm{MHz}$. Fig.~\ref{fig:supergaussian_envelope_amp_delta_max} shows the experimentally measured population of the first excited state $\ket{1}$ after applying the pulse to a qubit initialized in the ground state $\ket{0}$: a continuous plateau-like region with $P_1\approx1$ appears above $\Delta_{\rm max} \gtrapprox 4 \Omega_{2\pi}$ and $\Omega \gtrapprox 1.5\Omega_{2\pi}$.

\begin{figure}
		\centering

		\includegraphics[width=\defaultfigurewidth]{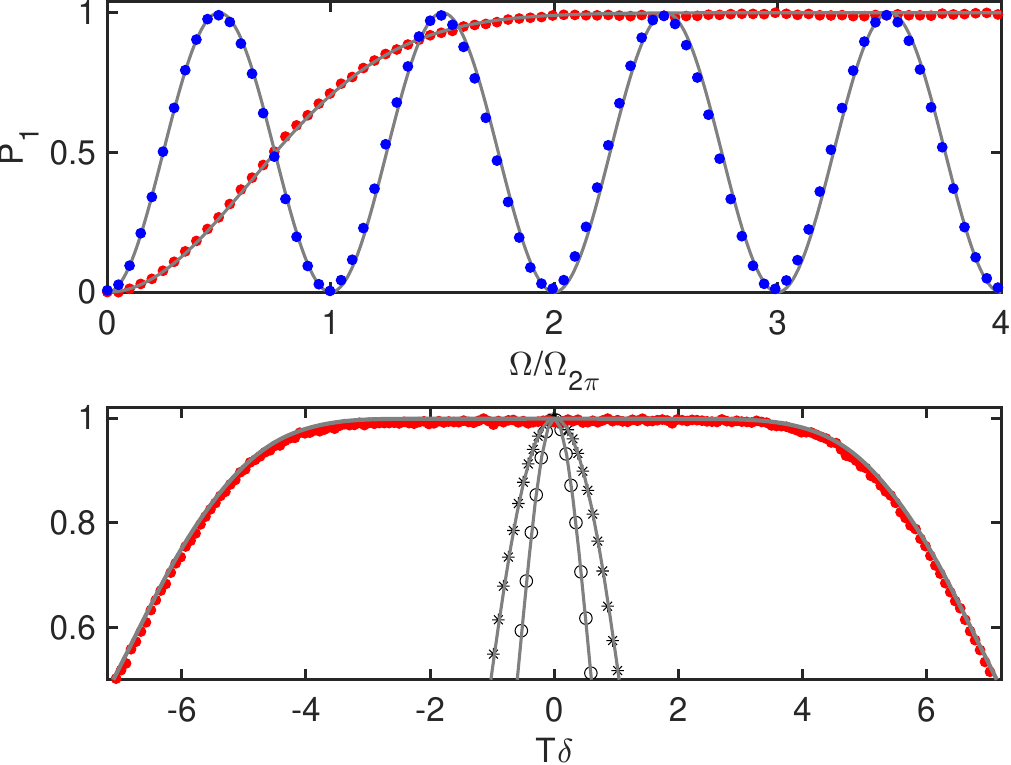}
		\caption{Top: the amplitude robustness of the super-Gaussian pulse (red dots), with $\Delta_{\rm max} =10.8 \Omega_{2\pi}$, compared to the usual non-modulated Rabi pulse (blue dots). Bottom: the detuning robustness of the super-Gaussian pulse (red dots) with $\Delta_{\rm max} =10.8 \Omega_{2\pi}$ and $\Omega= 2.5 \Omega_{2\pi}$. The gray circles and stars correspond to a Rabi $\pi$ ($\Omega = \Omega_{2\pi}/2$, $\Delta_{\rm max}=0$) and a $3\pi$ ($\Omega = 3\Omega_{2\pi}/2$, $\Delta_{\rm max}=0$) pulse respectively. The solid gray lines on both panels show the theoretical prediction with no free parameters.
		} 
		\label{fig:supergaussian_amp_and_freq}
	\end{figure}

	Fig.~\ref{fig:supergaussian_amp_and_freq} provides experimental and theoretical evidence that such a pulse ($\Delta_{max}=10.8\Omega_{2\pi}$) provides \textit{simultaneous} robustness in detuning ($\delta = \omega_q - \langle\omega_d\rangle$) and an amplitude with a wide margin for error: $P_1>99.9\%$ is prepared when $T\delta \lessapprox 3$ and $\Omega \gtrapprox 2.45\Omega_{2\pi}$.
	For comparison, the same holds true for a non-modulated $\pi$ pulse for deviations of less than only $\approx 0.02\Omega_{2\pi}$, in either the amplitude or detuning.
	
	In this regime, it is easy to develop an intuitive understanding of these effects: a small amplitude variation merely changes the on-resonance splitting of the instantaneous eigenstates, with no effect on the initial and final mixing angle. As $\Delta_{\rm max}$ increases, so does the rate of change of the eigenenergies; then a larger splitting $\Omega$ is necessary to avoid the Landau-Zener crossing, which gives rise to the triangular region in the upper left corner of Fig.~\ref{fig:supergaussian_amp_and_freq}, where the population is not transferred to the excited state. Likewise, a detuning which is small in comparison with $\Delta_{\rm max}$ (which can be made almost arbitrarily large), only shifts the moment at which the pulse is resonant with the qubit. Provided that it happens close to $t=0$, where $\Omega(t) \approx \Omega_0$, the splitting of the spectrum will be sufficient and will lead to population transfer. Therefore the detuning robustness grows with $\Delta_{\rm max}$, and is approximately $\Omega$ -- independent (provided that it is sufficiently large).

\begin{figure}
		\centering
		\includegraphics[width=\defaultfigurewidth]{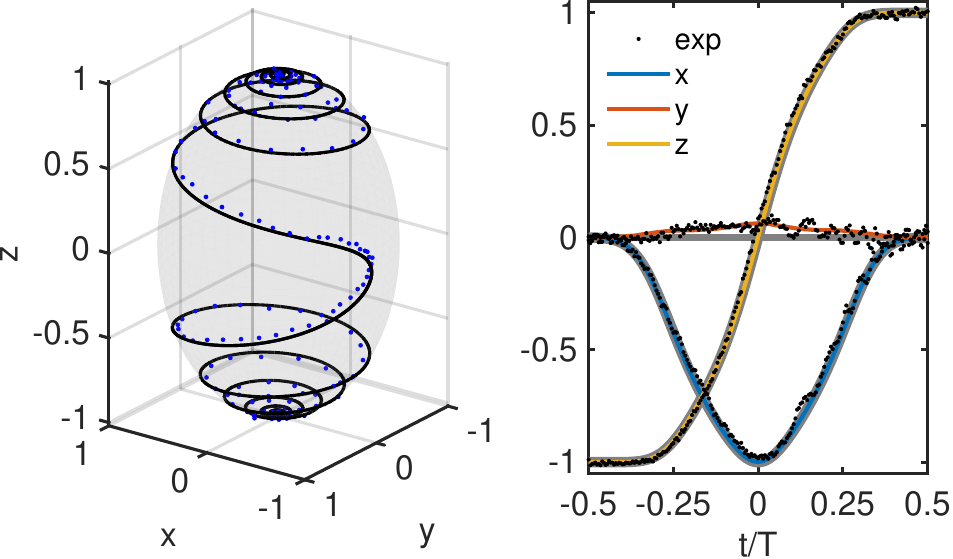}
		\caption{Left: the measured trajectory (dots) in the frame rotating at the qubit frequency, along with a theoretical model (solid line). Right: the trajectory in the frame co-rotating with the pulse. The theory is given by the solid colored lines, the adiabatic trajectory by the solid gray lines, while the experimental data is shown as dots. The pulse parameters were $\Omega=8\Omega_{2\pi}$, $\Delta_{\rm max}=40\Omega_{2\pi}$ and the pulse duration was $T=400~\mathrm{ns}$.
		}
		\label{fig:SG_tomo}
	\end{figure}

	The approximate adiabaticity of this protocol is demonstrated in Fig.~\ref{fig:SG_tomo}: the left panel shows the qubit trajectory (i.e. the components of the density matrix as a function of time) in the frame rotating at its frequency (in which one typically operates) on the Bloch sphere, while the right panel shows the trajectory in the rotating frame as defined above. 	
	The difference between the two frames is just a rotation around the $z$ axis by the angle $\phi(t) = \int_{-T/2}^{t} \Delta(\tau) \,d\tau$, and as $\Delta(t)$ is odd w.r.t $t$ they coincide at $t=\pm T/2$.

	The realized trajectory is close to the adiabatic one, the most prominent difference being a small non-zero $y$ component of the density matrix.
	
\begin{figure}
		\centering
		\includegraphics[width=\defaultfigurewidth]{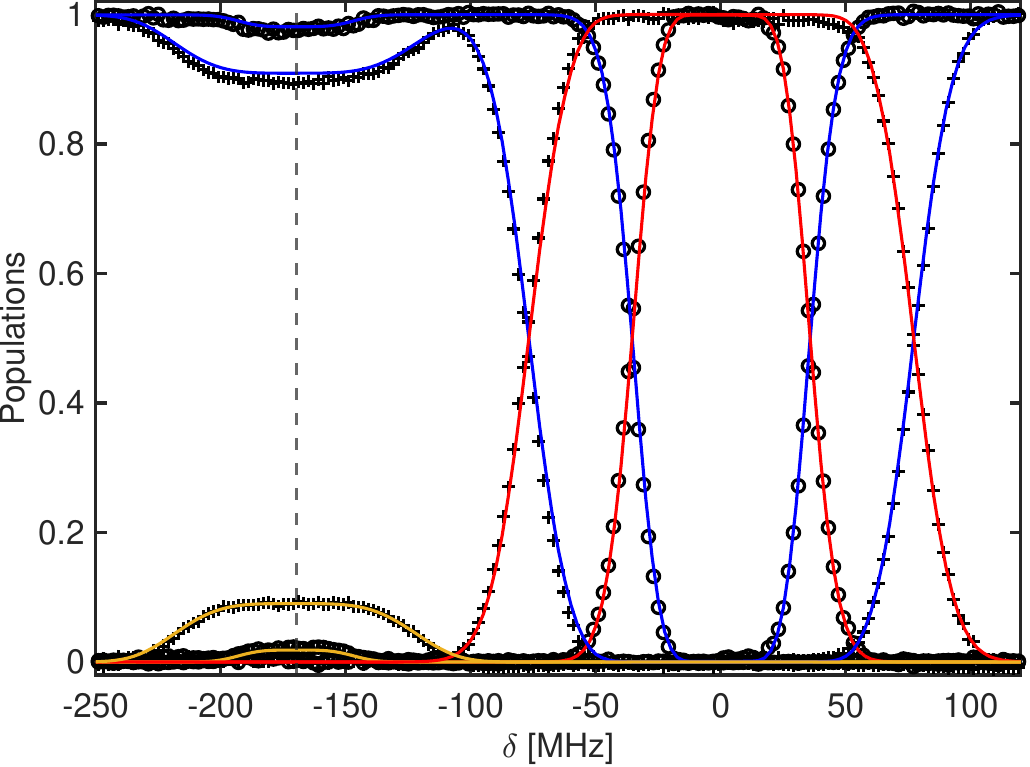}
		\caption{Theoretically predicted populations of the ground (blue), first (red) and second (yellow) excited states as a function of detuning with respect to the qubit frequency ($\delta=0$). The circles are experimental data for $\Omega=2.5\Omega_{2\pi}$ and $\Delta_{\rm max}=10.8\Omega_{2\pi}$ and the crosses for $\Omega=4.5\Omega_{2\pi}$ and $\Delta_{\rm max}=21.6\Omega_{2\pi}$. The pulse duration was $T=200\mathrm{ns}$. The vertical dashed line marks the $-E_c/2$ detuning from the qubit frequency.}
		\label{fig:sg_vs_detuning_3lvl}
	\end{figure}

	More detailed numerical analyses of the robustness and the adiabaticity of the protocol are given in appendices~\ref{app:SG} and~\ref{app:sg_vs_rect_env} respectively. Appendix~\ref{app:SG} also discusses the effects of the presence of the second excited state: this places a practical limit on the pulse duration $T$ and $\Delta_{\rm max}$. For realistic parameters, as the ones presented above, the cross-coupling effects are negligible. This is demonstrated in Fig.~\ref{fig:sg_vs_detuning_3lvl}: even with the large bandwidth of the pulse it is possible to selectively drive the transition between the ground and the first excited state, provided that the modulation depth $\Delta_{\rm max}$ is smaller than the separation between the resonant frequency and the 2-photon transition frequency $f_{02}=\frac{f_{01}+f_{12}}{2} \approx f_{01} - E_c/2$.

	\section{Arbitrary $X/Y$ Bloch sphere rotations} \label{rotation}	
	
	\begin{figure}
		\centering

		\includegraphics[width=\defaultfigurewidth]{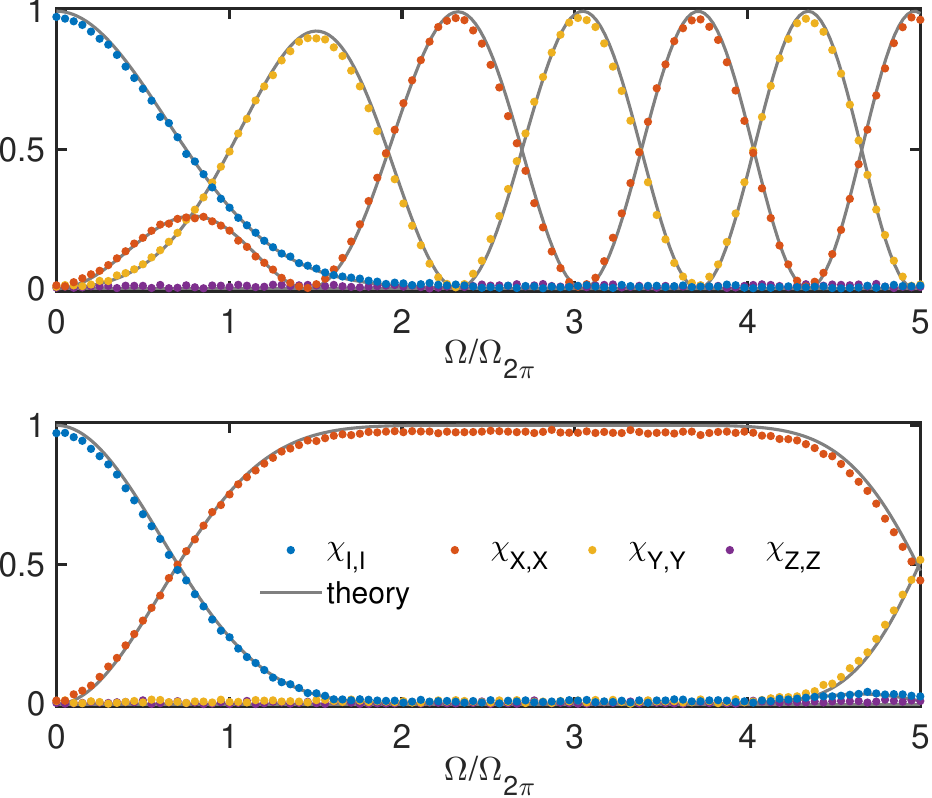}
		\caption{The diagonal elements of the process matrix $\chi$ for the super-Gaussian population transfer pulse (top) and the robust $\pi$ gate (bottom) as a function of the pulse amplitude $\Omega$. The solid lines show the theoretical prediction while the dots correspond to the experimental $\chi$ reconstruction.		
		}
		\label{fig:expectation_pauli_SG_vs_trig_pi}
	\end{figure}
	
	\begin{figure}
		\centering

		\includegraphics[width=\defaultfigurewidth]{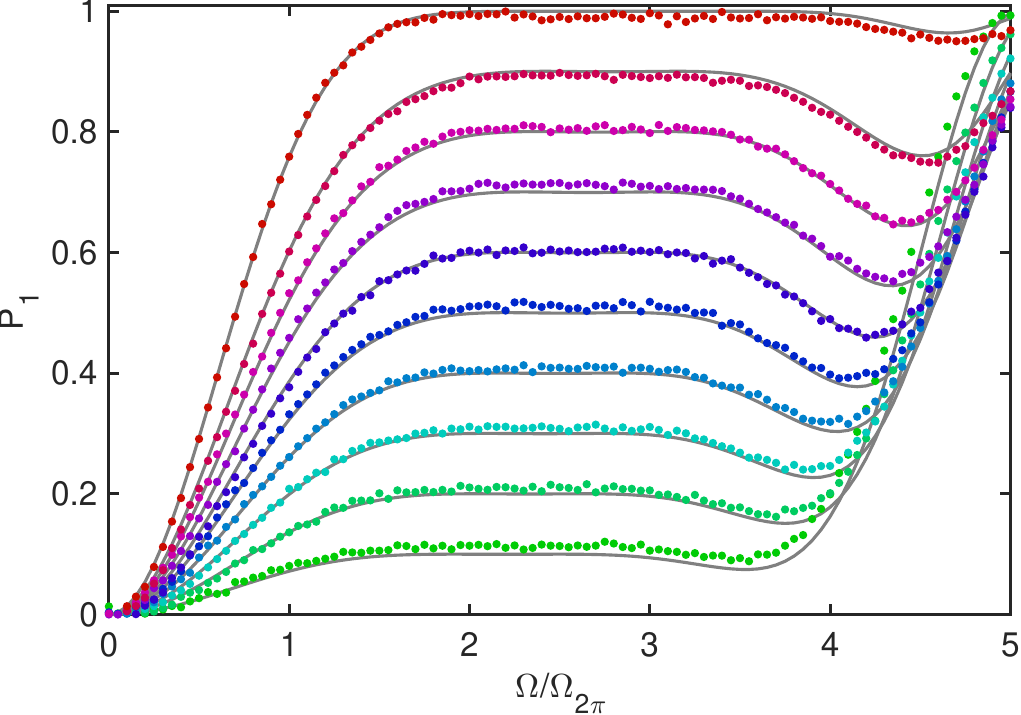}
		\caption{The experimentally measured population $P_1$ of the excited state (colored dots) after applying the pulses designed to obtain $P_1 = \{0.1,0.2,...,1\}$. The solid gray lines show the theoretical prediction with no free parameters.
		}
		\label{fig:EXP_P1_pulses_trigonometric}
	\end{figure}
	
	\begin{figure}
		\centering
		\includegraphics[width=\defaultfigurewidth]{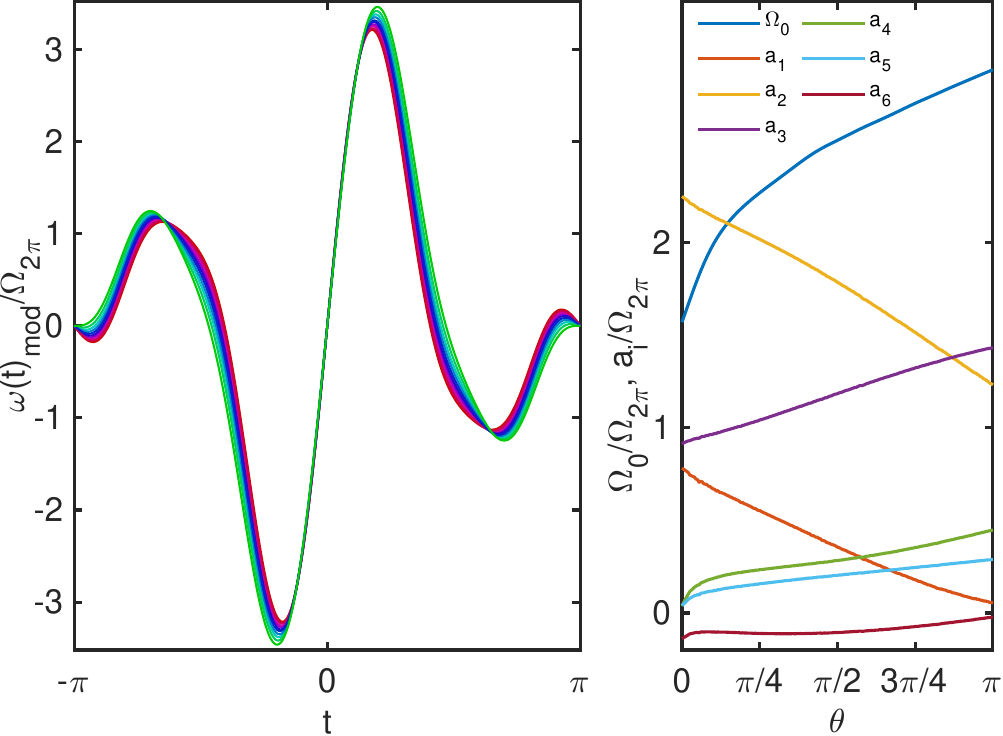}
		\caption{Left: the detuning necessary to realize the pulses shown in Fig.~\ref{fig:EXP_P1_pulses_trigonometric} (the traces are color-matched). Right: the pulse parameters, as defined in the text, as a function of the rotation angle. 
		}
		\label{fig:pulses_trigonometric}
	\end{figure}
	
	While the previous pulses enable robust population transfer between the ground and the excited state, they are not general $\pi$ rotations: nothing guarantees that the resulting unitary transformation $U$ has the proper phase factors - the state transfer experiment only demonstrates that $|U_{10}|^2 \approx 1$.  These pulses can be further characterized by performing quantum process tomography: for a unitary propagator $U = c_0 I - i \sum_{j=1}^3 c_j \sigma_j$ ($c_i \in \mathbb{R}$), the elements of the standard process matrix $\chi$ correspond to $\chi_{ii}=|c_i|^2$. For a $\pi$ rotation one has $c_\mathrm{I}=0$ ($\chi_{\mathrm{I,I}}=0$), and if the rotation is in the vertical plane (i.e. the pulse is resonant with the transition) $c_\mathrm{Z}=0$ ($\chi_{\mathrm{Z,Z}}=0$), while $|c_\mathrm{X}|^2 + |c_\mathrm{Y}|^2=1$. If the $\pi$ rotation is  generated solely by $\sigma_\mathrm{X}$ then $|c_\mathrm{X}|^2=\chi_{\mathrm{X,X}}=1$.

	The top panel of Fig.~\ref{fig:expectation_pauli_SG_vs_trig_pi} shows  the results of QPT for the population transfer pulse, as a function the pulse amplitude $\Omega$.	
	As $\chi_{\mathrm{I,I}} \approx 0$ it is indeed a $\pi$ rotation. However, the plane of rotation, while vertical ($\chi_{\mathrm{Z,Z}}\approx 0$), is amplitude dependent, as evidenced by the $\chi_{\mathrm{X,X}}$ and $\chi_{\mathrm{Y,Y}}$ components.

	To obtain a robust set of arbitrary angle $X/Y$ rotations an alternative approach was undertaken: A target unitary transformation $U_0$ is specified, through the rotation angles $\theta$ and $\phi$ on the Bloch sphere. Pulse parameters $(\Omega,\Delta(t))$ are numerically optimized (using standard gradient descent techniques) such that the Frobenius norm $\|U_0 - U(\Omega,\Delta(t))\|_{\rm F}$ is minimized.

	As the angle $\phi$ is experimentally controlled by the phase of the drive pulse, and a precise calibration of the qubit frequency is required to avoid unwanted $z$ rotations, we focus on the amplitude ($\Omega$) robustness.
	
	The pulses are parametrized by their optimal amplitude $\Omega_0$ (with a rectangular envelope), as well as the time-dependent frequency $\omega_d(t)$.
	With a pulse duration $T=2\pi$ and time $t \in [-\pi,\pi]$, $\omega$  was parametrized as $\omega_d(t) =\omega_q + \sum_{k\in \mathbb{N}} a_k \sin(k t)$, or equivalently $\Delta(t) = \sum_{k\in \mathbb{N}} a_k \left[\sin(k t) + \cos(k t)kt \right]$, with the Fourier coefficients $a_k$ as free parameters.
	Unlike for the population transfer, here the phase difference between the qubit frame and the pulse frame needs to be considered, as it effectively sets the phase of the rotation pulse. As was shown in the previous section, by restricting the detuning to be an odd function of time this difference vanishes, hence no $\cos(k t)$ terms appear in the expansion.
	
	The Fourier coefficients $a_k$ were then numerically optimized such that the resulting pulse is an amplitude robust pulse, i.e.~it minimizes $\langle\|U_0 - U(\Omega,\Delta)\|_{\rm F}\rangle_{\Omega}$ for $\Omega$ close to $\Omega_0$. A $\pi$ pulse obtained in this manner is also shown in Fig.~\ref{fig:expectation_pauli_SG_vs_trig_pi}: it also exhibits amplitude robustness, but unlike the population-transfer pulse the plane of rotation is constant (with $\chi_{\mathrm{X,X}} \approx 1$ and $\chi_{\mathrm{I,I}},\chi_{\mathrm{Y,Y}},\chi_{\mathrm{Z,Z}} \approx 0$).

	Pulse parameters that implement rotations for different $\theta$'s can be found by this method:	Fig.~\ref{fig:EXP_P1_pulses_trigonometric}, shows a series of amplitude robust pulses, engineered such that $P_1=\sin(\theta/2)^2$; for a qubit initialized in $\ket{0}$, values are linearly spaced from 1 to 0.1. For all angles we observe a flat, plateau-like region, for which the desired $\theta$ rotation is implemented. Fig.~\ref{fig:pulses_trigonometric} shows the corresponding pulses, as well as their parameters.  Moreover, the control parameter values are a smooth function of $\theta$, allowing for an arbitrary angle rotation by interpolation of the parameter values (demonstrated in appendix \ref{app:arb_rot_int}).
	Even though robustness against frequency errors was not an optimization goal for these pulses they also offer better performance in this regard compared to the usual Rabi pulses, see section \ref{app:trig_det_rob} of the appendix.

\begin{widetext}
		
		\begin{figure*}
			\centering
			\includegraphics[width=\textwidth]{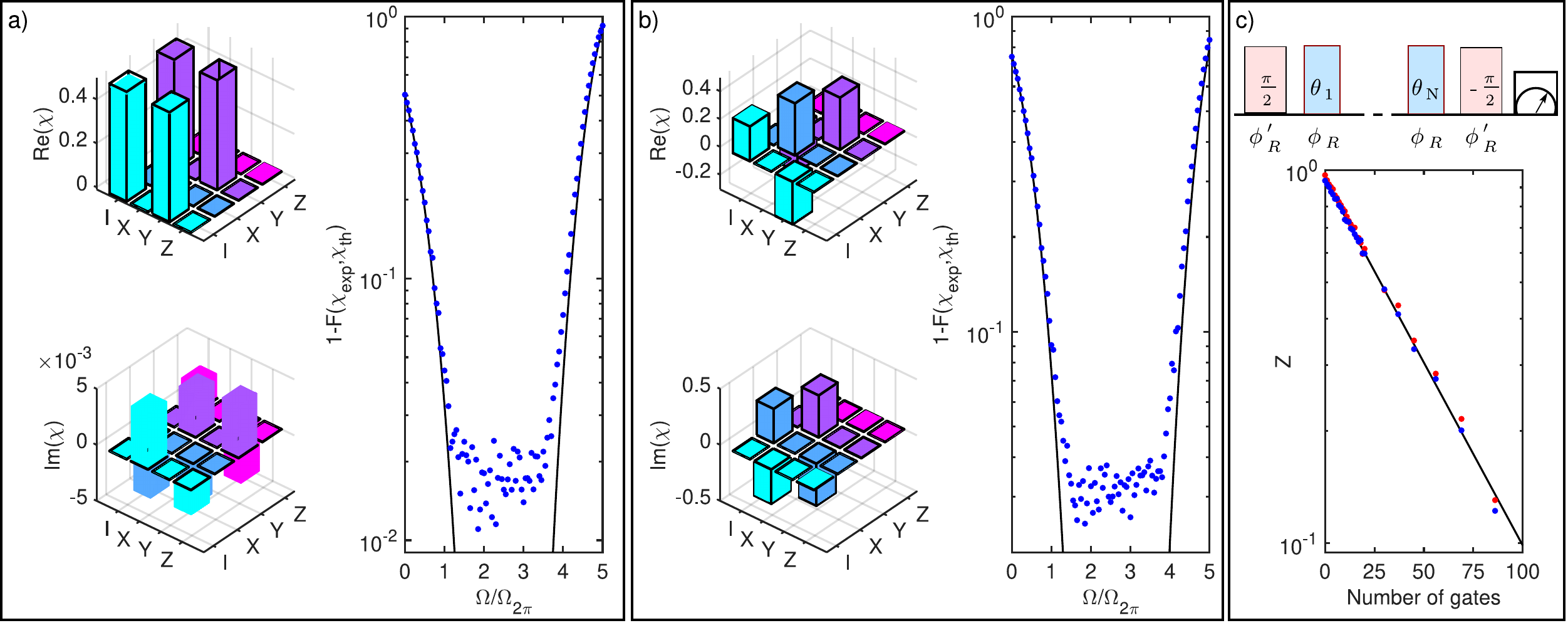}
			\caption{
				a) Real Re$(\chi(\frac{\pi}{2})_{y})$ and imaginary 
				Im$(\chi(\frac{\pi}{2})_{y})$ parts of the process matrix 
				for a $\left(\pi/2\right)_y$ rotation are shown in a cityscale representation, alongside with the variation of fidelity between the 
				theoretical and experimental process matrices for different values of the Rabi 
				coupling (same horizontal scale as that of Fig.~\ref{fig:EXP_P1_pulses_trigonometric}). 
				b) Real Re$(\chi(\theta)_{\hat{n}})$ 
				and imaginary Im$(\chi(\theta)_{\hat{n}})$ parts of the process 
				matrix for an arbitrary $x,y-$rotation ($\theta_{\hat{n}}$) of a single qubit, 
				initialized in state $|0\rangle$ are shown in a cityscape representation. 
				The angle of rotation, $\theta=\frac{2\pi}{3}$, 
				is chosen arbitrarily (and corresponds to $P_1=0.75$) and the axis of rotation 
				is chosen to be aligned at an angle of $\phi=-\frac{\pi}{4}$ with $y-$axis, such that the
				generator of the  rotation is 
				$\sigma_{\hat{n}}=\cos(\phi) \sigma_y + \sin(\phi) \sigma_x$. The variation of fidelity between the 
				theoretical and experimental process matrices is shown to the right.
				In both cases, experimental process matrices 
				correspond to the centre of the respective flat regions.
				The experimentally-obtained elements of the process matrices are shown with colored opaque bars, while the black wire-frame bars correspond to the theoretically expected process matrices.
				c) Upper panel: the pulse sequence for the randomized benchmarking where $\phi_{R}'=\phi_R+\pi/2$, lower panel: experimental results from randomized benchmarking of the
				amplitude-robust quantum gates (blue dots), compared to the normal Rabi pulses of the same duration (red dots), along with an exponential decay fit (solid black line).
			}
			\label{fig:qpt_reconstruction}
		\end{figure*}		
	\end{widetext}

	In order to study the efficacy and robustness against amplitude variations, we perform QPT for amplitudes $\Omega \in [0, 5 \Omega_{2\pi}]$, and quantify the fidelity of these gates as $F = {\rm Tr}(\chi . \chi_{\rm exp})$, where $\chi$ is the process matrix for the desired operation and $\chi_{\rm exp}$ the experimentally reconstructed process matrix.
	Fig.~\ref{fig:qpt_reconstruction} 	shows the variation of the fidelity versus the Rabi coupling for two arbitrarily chosen $(\theta,\phi)$ angles: the fidelity stays high ($\approx0.98$) for a wide enough range of amplitudes, while it is theoretically expected to reach $\approx 1-10^{-6}$. This flat region of high fidelity corresponds to the flat region of the populations, shown in Fig.~\ref{fig:EXP_P1_pulses_trigonometric}. Experimentally reconstructed process matrices at the center of the plateau, also shown in Fig.~\ref{fig:qpt_reconstruction}, are in close agreement with the theoretically expected ones. 

	Further, to characterize the overall performance of these amplitude-robust pulses, we additionally performed randomized benchmarking. This also allows us to demonstrate that the limiting factor for the experimentally observed fidelity is related to the sample and not to the pulses themselves.
	
	We consider the following sequence of operations consisting of $\rm{N}+2$ gates given by: $S=(\pi/2)_{\phi_R+\pi/2} (\theta_{\rm N})_{\phi_R} \dots (\theta_1)_{\phi_R}  (-\pi/2)_{\phi_R+\pi/2}$, where the axis of rotation, set by $\phi_R \in \left[ -\pi/2, \pi/2\right]$, is chosen randomly and $\theta_i$'s ($i \in [1,{\rm N}]$) are the randomly chosen angles of rotation. 	This way the whole sequence is effectively a diagonal operator, which corresponds to a rotation generated by $\sigma_z$. Therefore, for a qubit initialized in the ground state we should have $|\langle 0 | \rho_f |0\rangle|^2=1$.	
	
	However, in reality, decoherence and the non-ideality of the gate implementation will lead to a finite population of the excited state, which will tend to $0.5$ with an increasing number of operations (i.e. the qubit will end up in a maximally mixed state). Therefore we take the population of the ground state as a measure of the overall circuit fidelity.
	
	The experimentally obtained curve presenting exponentially decaying fidelity with increasing number of gates is shown in panel c) of Fig.~\ref{fig:qpt_reconstruction}: the loss per gate  is found to be $2.87\%$.
	For comparison we performed the same procedure with standard (non-modulated) Rabi pulses, obtaining the same $\approx 3\%$ error per gate, set by the pulse duration (here $T=120~\mathrm{ns}$) and the relatively short coherence times. Note however that these Rabi pulses have to be very well calibrated and they do not
	have the robustness property with respect to amplitude,	shown in panels a) and b) of Fig.~\ref{fig:qpt_reconstruction}, of phase-modulated pulses.	

	Therefore, even higher fidelities can be achieved in principle with higher-quality samples, where the errors due to instrumentation and calibration start to be comparable with those resulting from decoherence.

	\bigskip
	
	\section{Discussion and Conclusions} \label{discussion}	
	
	For the success of a quantum-computing research programme with superconducting qubits, reducing the sensitivity to imperfections and noise is of utmost importance. With gradual improvements in the decoherence times over the last years, errors in amplitude and frequency of the pulses used to manipulate these systems become the dominant source of fidelity loss.
	
	Here we have proposed theoretically and realized experimentally control protocols for robust qubit state manipulation employing phase-modulated pulses. The control waveforms employed here are stable - meaning that errors in the operations do not increase rapidly under small perturbations of the control parameters. The number of optimization parameters is not very large, which makes the numerical optimization less time consuming. In the first part we demonstrated that a simple scheme, based on a Landau-Zener-like process, can result in an operation which realizes a population transfer between two levels with considerable robustness: any amplitude above a threshold value leads to a complete population transfer, while the frequency of the pulse can be detuned by several tens of $\mathrm{MHz}$ from the transition frequency.
	In the second part we presented a set of pulses which realize amplitude robust $X/Y$ rotations. The performance of these pulses was evaluated using quantum process tomography as well as randomized benchmarking, showing that they realize the desired operation with high fidelity. These methods can be readily applied to any other experimental platform (trapped ions, NV centers, etc.) where a two-level system is manipulated using microwave or laser pulses. Furthermore, it should be possible to generalize this approach for multi-state (e.g. qutrit) and multi-qubit control.

	\acknowledgments	
	
	We are grateful to Aidar Sultanov for help with the experiments at various stages of this work. This project has received funding from the European Union project OpenSuperQ+. We acknowledge also financial support from Grant No. FQXi-IAF19-06 (“Exploring the fundamental limits set by thermodynamics in the quantum regime”) of the Foundational Questions Institute Fund (FQXi), a donor advised fund of the Silicon Valley Community Foundation, as well as from MEC Global funding. 
	This work was performed as part of the Academy of Finland Centre of Excellence program (project 352925).

	\FloatBarrier
	
	\bibliography{Bibliography_robustpulses}
	
	\clearpage
	
	\appendix
	
	\FloatBarrier
	
	\section{Robust population transfer with constant amplitude pulses}
	
	Here we discuss the case of LZMS drive realized with constant amplitude pulses. The scheme is principally the same as the one in the main text, where the detuning $\Delta(t)$ is a linear function, but the pulses are rectangular.
	
\begin{figure}
		\centering
		\includegraphics[width=\defaultfigurewidth]{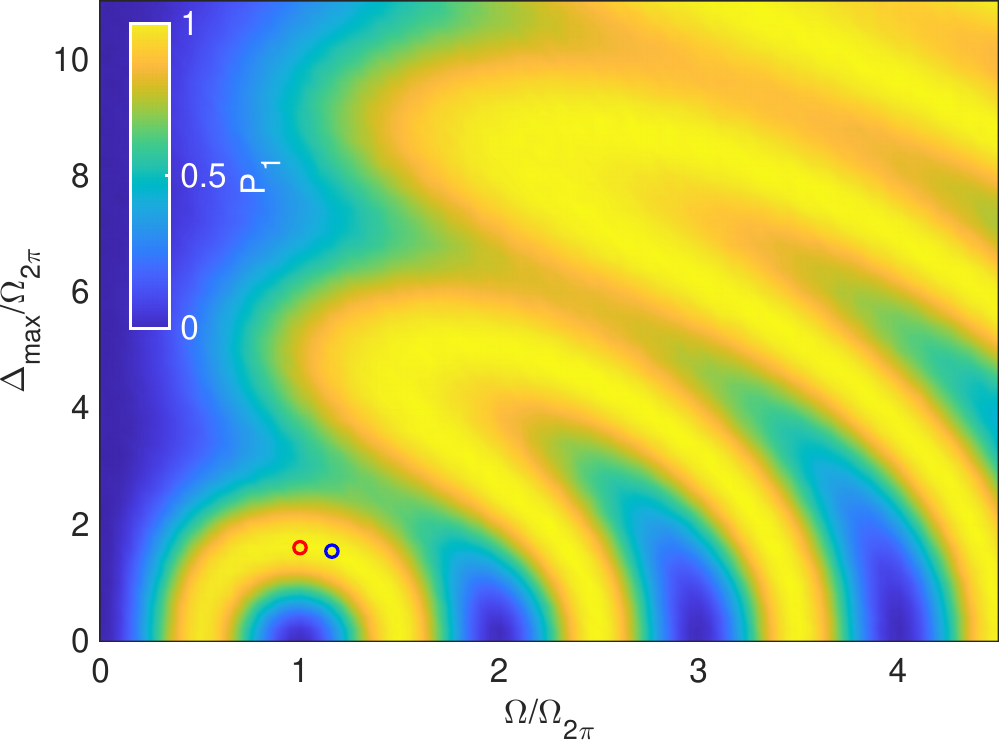}
		\caption{Experimentally obtained population of $\ket{1}$, $P_1$ by applying the $(\Omega,\Delta_{max})$ pulse to the qubit initialized in $\ket{0}$. The red and blue circles indicate an amplitude or detuning robust pulse respectively.}
		\label{fig:rectangular_envelope_amp_delta_max}
	\end{figure}

	Fig.~\ref{fig:rectangular_envelope_amp_delta_max} shows the experimentally measured population of the first excited  state $\ket{1}$ after applying such a pulse to a qubit initialized in the ground state $\ket{0}$.
	As opposed to the plateau obtained with super-Gaussian pulses (see Fig.~\ref{fig:supergaussian_envelope_amp_delta_max}), several non-connected stripes with $P_1=1$ are observed for rectangular-envelope pulses, appearing at successively higher Rabi and detuning frequencies. Still, we can identify two points on the first lobe of this pattern, which realize pulses robust to amplitude offsets (red circle in Fig.~\ref{fig:rectangular_envelope_amp_delta_max}), and to detuning offsets (blue circle in Fig.~\ref{fig:rectangular_envelope_amp_delta_max}) respectively. 
	
\begin{figure}
		\centering
		\includegraphics[width=\defaultfigurewidth]{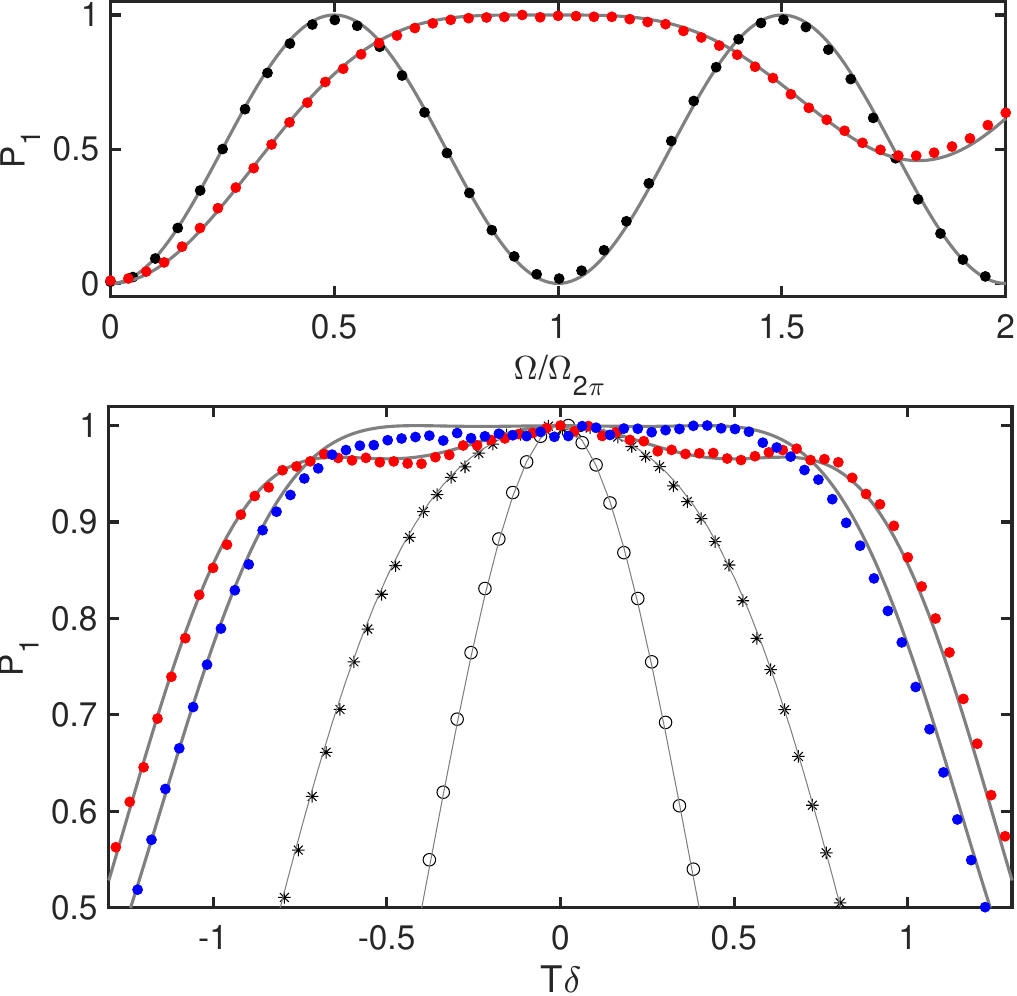}
		\caption{Top: the amplitude robust pulse (red circle on Fig.~\ref{fig:rectangular_envelope_amp_delta_max}, $\Delta_{\rm max}=1.64\Omega_{2\pi}$, $\Omega=\Omega_{2\pi}$) compared to the resonant Rabi drive. Bottom: the detuning robustness of the amplitude robust pulse and the detuning robust one (blue circle in Fig.~\ref{fig:rectangular_envelope_amp_delta_max}, $\Delta_{\rm max}=1.54\Omega_{2\pi}$ $\Omega = 1.16\Omega_{2\pi}$).
			The gray circles and stars correspond to a Rabi $\pi$ and a $3\pi$ pulse respectively. The solid gray lines on both panels show theoretical predictions with no free parameters.
		}
		\label{fig:rectangular_1st_lobe_amp_and_freq}
	\end{figure}

	The top panel of Fig.~\ref{fig:rectangular_1st_lobe_amp_and_freq} highlights the amplitude robustness in comparison with a resonant Rabi pulse: $P_1>99.9\%$ is achieved for the range of amplitudes $\Omega \in [0.82\Omega_{2\pi}, 1.12\Omega_{2\pi}]$. 
	
	At the same time this pulse is not robust against detuning from the qubit frequency ($\delta = \omega_q - \omega_d$), as shown in in the bottom panel of Fig.~\ref{fig:rectangular_1st_lobe_amp_and_freq}. A pulse with slightly different parameters (blue circle in Fig.~\ref{fig:rectangular_envelope_amp_delta_max}), on the other hand, results in a population transfer with $P_1>99.9\%$ when it is detuned from the qubit frequency by less than  $T\delta \lessapprox 0.5$.
	
	The performance of these rectangular pulses is quite similar to those found in \cite{ding2020breaking,ai2021experimentally}, as the detuning profile obtained there is almost linear, with the added benefit of a simpler implementation.

	While the higher order lobes shown in Fig.~\ref{fig:rectangular_envelope_amp_delta_max}, can in principle offer greater amplitude/frequency robustness at the expense of a larger Rabi and modulation frequencies, they are not of great interest due to the higher levels of power required. Instead we focus here on the theoretical understanding of the pattern of fringes observed.
	
	The Hamiltonian given by Eq.~(\ref{eq:hamiltionian}) with $\Omega$ time-independent has two instantaneous eigenstates $\ket{E_{+}(t)}$ and $\ket{E_{-}(t)}$ with energies $E_{\pm}(t)=\pm \frac{\hbar}{2}\sqrt{\Omega^2 + \Delta(t)^2}$.
	The eigenstates can be parametrized as $\ket{E_{-}(t)}= (-\sin(\Theta), \cos(\Theta))^\mathrm{T}$ and $\ket{E_{+}(t)}= (\cos(\Theta),\sin(\Theta))^\mathrm{T}$, where the mixing angle is given by $\tan(\Theta(t)) = \frac{\Delta(t) + \sqrt{\Omega^2 + \Delta^2(t)}}{\Omega}$. For $\Delta(t) = t \dfrac{2\Delta_{\rm max}}{T}$ we have $\Delta(\pm\frac{T}{2}) = \mp \Delta_{\rm max}$, and as a consequence the mixing angles then satisfy $\Theta(-\frac{T}{2}) + \Theta(\frac{T}{2}) = \frac{\pi}{2}$.
	
	If we denote the mixing angle at $t=-\frac{T}{2}$ by $\Theta_0$, then $\ket{E_{-}(-\frac{T}{2})}= (-\sin(\Theta_0),\cos(\Theta_0))^\mathrm{T}$ and $\ket{E_{+}(-\frac{T}{2})}= (\cos(\Theta_0),\sin(\Theta_0))^\mathrm{T}$, as well as 
	$\ket{E_{-}(\frac{T}{2})}= (-\cos(\Theta_0),\sin(\Theta_0))^\mathrm{T}$ and $\ket{E_{+}(\frac{T}{2})}= (\sin(\Theta_0),\cos(\Theta_0))^\mathrm{T}$. 
	
	Any state $\ket{\psi}$ can be decomposed into $\ket{E_{\pm}}$ as $\ket{\psi} = c_{-}\ket{E_{-}} + c_{+}\ket{E_{+}}$ and the time evolution is given by the time dependence of $c_\pm(t)$. 	
	For a qubit initialized in the ground state $\ket{0}=(0,1)^\mathrm{T}$ initially $c_{-}(-\frac{T}{2})=\cos(\Theta_0)$ and $c_{-}(-\frac{T}{2})=\sin(\Theta_0)$, and Rabi flopping will occur.
	For an adiabatic trajectory $|c_\pm|=\mathrm{const}$, with only a time-dependent phase $c_{\pm}(t) = c_{\pm}(-\frac{T}{2}) e^{\pm i\varphi(t)/2}$, where $\varphi(t) = \int_{-\frac{T}{2}}^{t} \sqrt{\Omega^2 + \Delta^2(t)} \,dt$.
	Only if $\varphi(\frac{T}{2}) = (2k+1)\pi$ ($k \in \mathbb{Z}$) will the qubit end up in the excited state $\ket{1}=(0,1)^T$.
	
\begin{figure}
		\centering
		\includegraphics[width=\defaultfigurewidth]{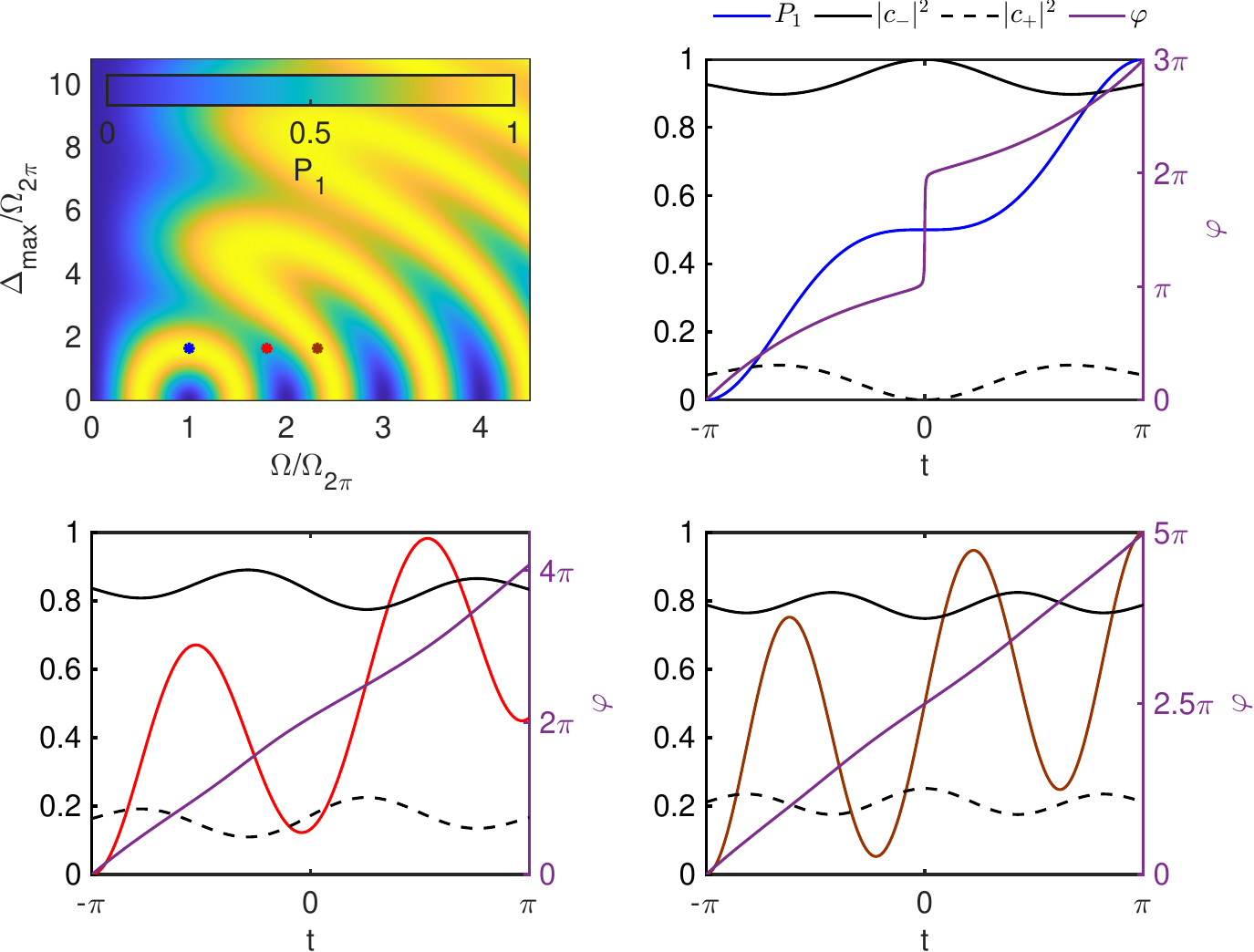}
		\caption{Top left: simulation of the population tansfer as a function of $\Omega$ and $\Delta_{\rm max}$. Other panels: the time evolution, given by the population of the excited state $P_1$ (colored), coefficients $|c_{\pm}|^2$ (black, solid and dashed respectively) and their relative phase $\varphi$ (purple, right scale) for the 3 points highlighted in the top-left panel (the colors of the dots and $P_1$ traces are matched).
		}
		\label{fig:fringes_simulation}
	\end{figure}

	Even if the trajectory is not adiabatic the same holds true: the final state $\ket{1}$ is obtained only when $|c_{\pm}(-\frac{T}{2})| = |c_{\pm}(\frac{T}{2})|$ with a relative phase factor of $e^{i\varphi} = -1$.
	This condition leads to the interference-like pattern seen in Fig.~\ref{fig:rectangular_envelope_amp_delta_max}, reproduced here numerically in Fig.~\ref{fig:fringes_simulation}.
	The same figure shows several trajectories, for different parameter values, demonstrating this interference effect: in all cases the trajectory is approximately adiabatic ($|c_{\pm}|\approx {\rm const}$) and $|c_\pm(-\frac{T}{2})| = |c_\pm(\frac{T}{2})|$ is satisfied, only the accumulated phase $\varphi$ is different: $\varphi(\frac{T}{2})= 3\pi$ and $\varphi(\frac{T}{2})= 5\pi$ leads to $P_1(\frac{T}{2})= 1$, while for $\varphi(\frac{T}{2})= 2\pi$ we have $P_1(\frac{T}{2})\approx 0.5$.
	
	Unlike with a constant amplitude pulse, the super-Gaussian envelope ensures that the mixing angle at $t=\pm\frac{T}{2}$ is $\approx0$ and $\approx\frac{\pi}{2}$ respectively. Therefore there will be no Rabi flopping, and the relative phase of the two eigenstates does not matter, leading to the plateau observed in Fig. \ref{fig:supergaussian_envelope_amp_delta_max} of the main text and Fig. \ref{fig:sg_robustnes_amp_slope}.

	\section{Robustness of linearly modulated super-Gaussian pulses} \label{app:SG}
	
	In the rotating wave approximation the Hamiltonian of a three-level system subject to a drive can be written as:
	
	\begin{equation}
		H = \hbar\begin{pmatrix}
			-\Delta_{01}(t) & \Omega_{01}(t)/2 & 0 \\
			\Omega_{01}(t)/2& 0 &	\Omega_{12}(t)/2\\
			0 & \Omega_{12}(t)/2 & \Delta_{12}(t)
		\end{pmatrix},
		\label{eq:hamiltionian3lvl}
	\end{equation}
	where $\Delta_{01}$ and $\Delta_{12}$ are the detunings between the drive frequency and the $f_{01}$ and $f_{12}$ transition frequencies. $\Omega_{01}$ and $\Omega_{12}$ are the Rabi frequencies that drive the $01$ and $12$ transitions, differing only by the ratio of the transition dipole moments $\frac{\Omega_{12}}{\Omega_{01}}=\frac{g_{12}}{g_{01}}=\lambda$. We assume that the $02$ dipole moment $g_{02}=0$. The model presented here can reproduce the two-photon (i.e. Raman) transitions as well, and should describe well all of the effects that one typically observes with short duration pulses and/or high amplitudes.
	
	For a transmon we have $E_J/E_C \to \infty$ then $\lambda \to -\sqrt{2}$ and $f_{12} \to f_{01} - E_C$. Here we assume $\lambda=-\sqrt{2}$, as it represents the worst case scenario for the effects of the cross-coupling. We take the experimental value of $E_C=340~\mathrm{MHz}$.
	
\begin{figure}
		\centering
		\includegraphics[width=\defaultfigurewidth]{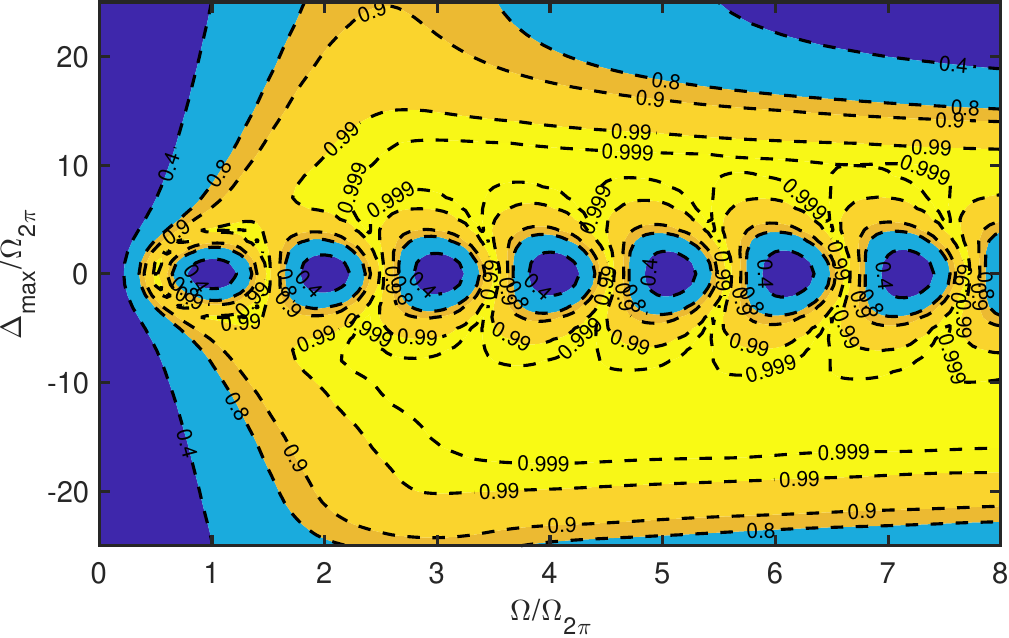}
		\caption{A contour plot of $P_1$ as a function of $(\Omega,\Delta_{\rm max})$ with a super-Gaussian envelope, simulated using Eq.~(\ref{eq:hamiltionian3lvl}).
		}
		\label{fig:sg_robustnes_amp_slope}
	\end{figure}

	A $T=60~\mathrm{ns}$ pulse with a super-Gaussian envelope was simulated using Eq.~(\ref{eq:hamiltionian3lvl}).
	Fig.~\ref{fig:sg_robustnes_amp_slope} shows the population of the first excited state $P_1$ as a function of $(\Omega,\Delta_{\rm max})$, and is comparable to Fig.~\ref{fig:supergaussian_envelope_amp_delta_max} of the main text.
	While the general structure is the same now it is clear that the protocol breaks down for large values of $\Delta_{\rm max}$, by driving either the 12 or 02 transitions. The model presented in the main text is invariant under $\Delta_{\rm max} \rightarrow -\Delta_{\rm max}$, as it just changes whether the qubit follows the lower or higher eigenstate of the Hamiltonian. Here, however, the presence of the second level breaks this symmetry and gives rise to the behavior  seen in Fig.~\ref{fig:sg_robustnes_amp_slope}.
	
\begin{figure}
		\centering
		\includegraphics[width=0.5\textwidth]{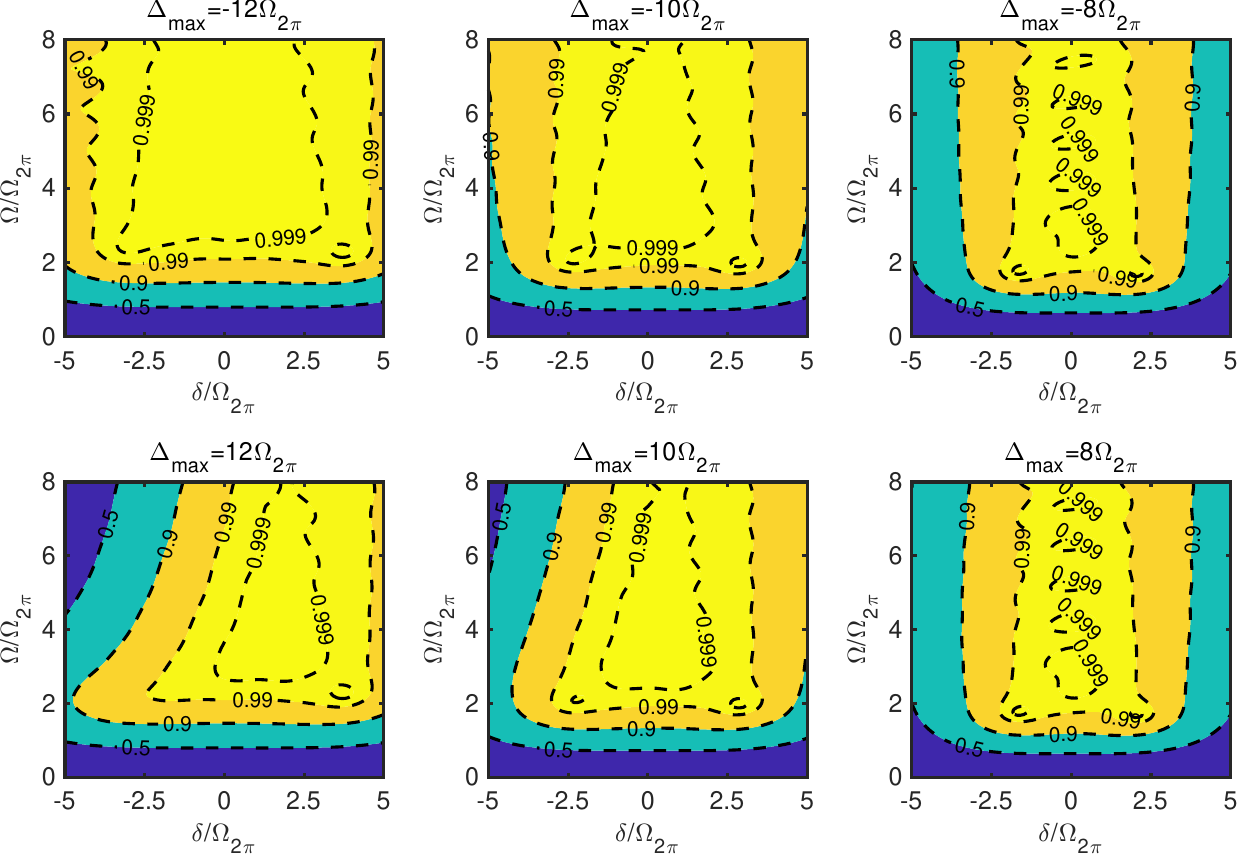}
		\caption{A contour plot of $P_1$, showing the detuning robustness of the protocol as a function of $\Omega$, for several values of $\Delta_{\rm max}$.
		}
		\label{fig:sg_robustnes_amp_delta}
	\end{figure}

	The detuning robustness can be studied in the same way; $(\delta,\Omega)$ maps are shown in Fig.~ \ref{fig:sg_robustnes_amp_delta}. The results are in accordance with the statements made about the detuning robustness in the main text: generally it grows with $\Delta_{\rm max}$ and is approximately independent of $\Omega$, above a threshold value. The existence of the second excited state shows up as in Fig.~\ref{fig:sg_robustnes_amp_slope}: it breaks the $\Delta_{\rm max} \rightarrow -\Delta_{\rm max}$ and $\delta \rightarrow -\delta$ symmetries.
	
	For longer pulses ($E_{C}T\ll1$) the detuning robustness, as well as the pulse bandwidth, is determined primarily by the modulation depth $\Delta_{\rm max}$: if it is sufficiently smaller than the anharmonicity one is still able to selectively drive the transition from the ground to the first excited state, without populating the second excited state. Fig.~\ref{fig:sg_vs_detuning_3lvl} in the main text demonstrates this theoretically, as well as experimentally: for a $T=200~\mathrm{ns}$ pulse one can find parameters $(\Omega,\Delta_{\rm max})$ that robustly drive the transmon to the first excited state; the second excited state is only populated if the pulse is detuned by $\delta \approx E_C/2$.
	
	Overall, one can conclude that for moderate values of $\Delta_{\rm max}$ (i.e $\Delta_{\rm max}\ll E_C T$) the protocol is unaffected by the existence of the second excited state. Additionally if one considers longer pulse durations higher order states can be disregarded, while still providing a considerable degree of robustness.

	\section{Super-Gaussian vs other envelopes}\label{app:sg_vs_rect_env}
	
	In the main text several criteria for the envelope were laid out, most important of which is the vanishing amplitude at the end of the pulse. The super-Gaussian envlope was chosen as a balance between the spectral width of the pulse and a lower peak amplitude compared to e.g.~a Gaussian pulse. An added benefit is that the flatness of the envelope near $t=0$ leads to an increased detuning robustness, which was not extensively studied for other RAP-like pulses. For a pulse detuned by $\delta$ from the transition frequency the eigenenergy spectrum will attain its minimum value at $t=\frac{T\delta}{2\delta_{max}}\neq 0$.
	
	The rapid-adiabatic-passage family of pulses offers a lot of possibilities, without a clear winner in terms of performance. Many of the used envelopes (e.g. a Gaussian, $\mathrm{sech}(t/\tau$, ...) have the same qualitative shape: a peak at $t=0$ with a relatively rapid drop-off. Here we make a comparison with a Gaussian envelope, and the findings should carry over to other envelopes.

\begin{figure}
		\centering
		\includegraphics[width=0.5\textwidth]{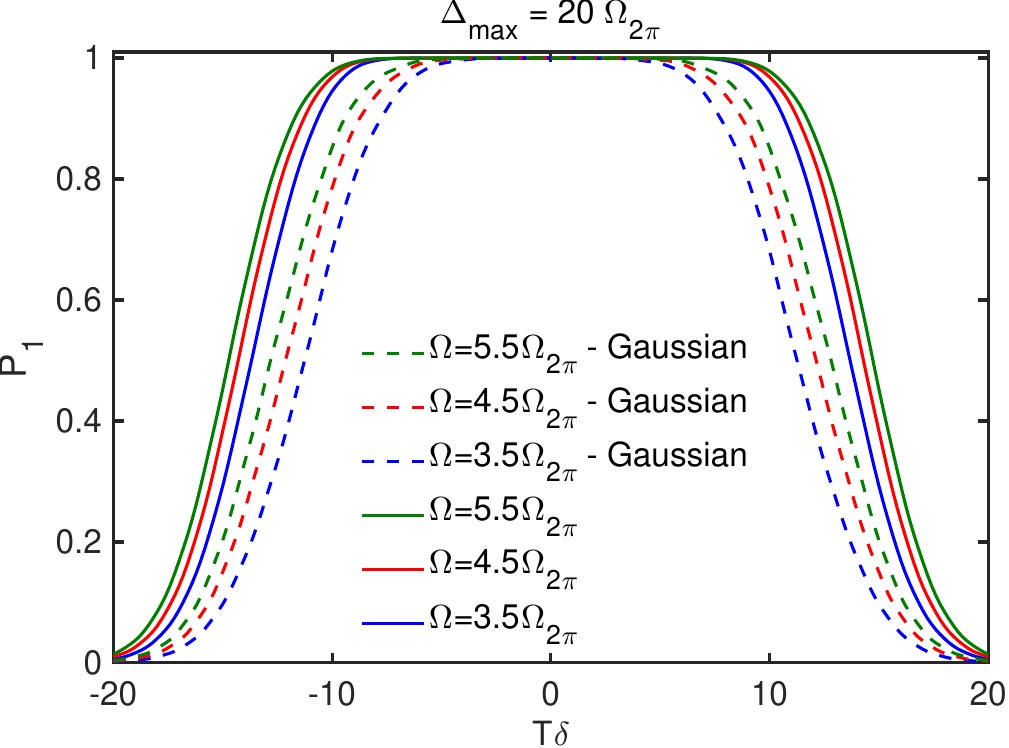}
		\caption{The detuning robustness of the Gaussian (dashed lines) and super-Gaussian (solid lines) as a function of the pulse amplitude, normalized to the same area-under-the-curve, for $\Delta_{\rm max}=20\Omega_{2\pi}$. 	
		}
		\label{fig:gaussian_vs_supergaussian_amp}
	\end{figure}

\begin{figure}
		\centering
		\includegraphics[width=0.5\textwidth]{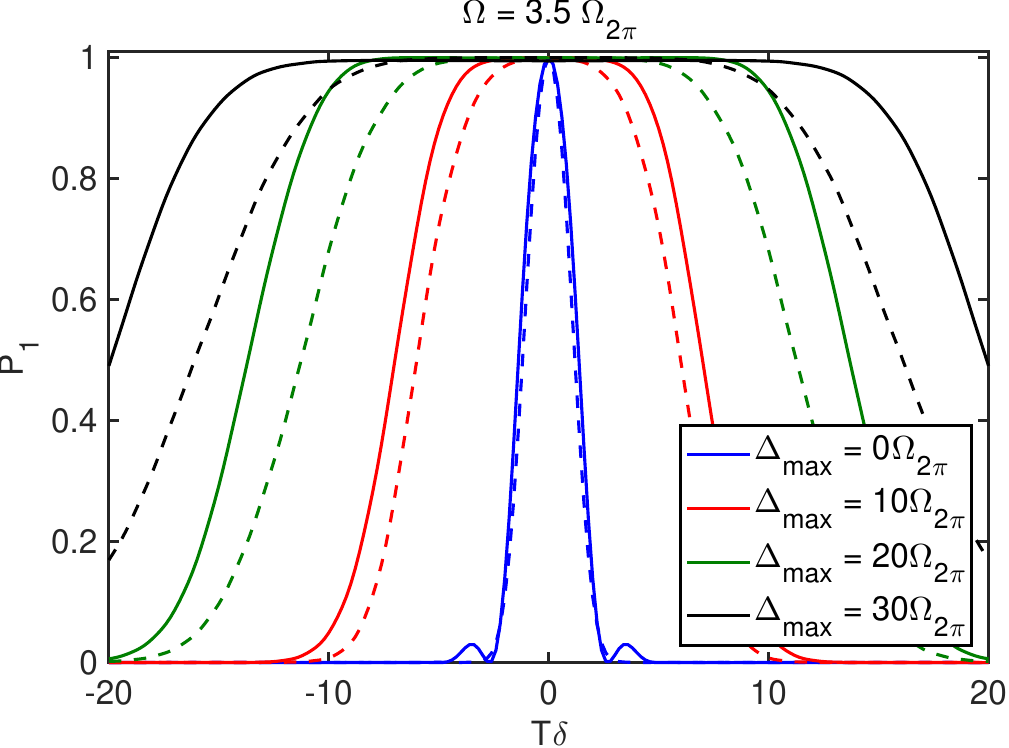}
		\caption{The detuning robustness of the Gaussian (dashed lines) and super-Gaussian (solid lines) as a function of the modulation depth $\Delta_{\rm max}$. The envelopes were normalized to the same area, with $<\Omega> = 3.5\Omega_{2\pi}$.
		}
		\label{fig:gaussian_vs_supergaussian_delta_max}
	\end{figure}

	For a Gaussian envelope the amplitude will be suppressed more compared to its peak value $\Omega(t=0)$ than for the super-Gaussian envelope. However, due to the envelope shape the Gaussian pulse has a higher peak amplitude. These are two competing effects, and it is not possible to immediately say which one is dominant. To investigate this we performed a numerical study of the detuning robustness: Figs.~\ref{fig:gaussian_vs_supergaussian_amp} and~\ref{fig:gaussian_vs_supergaussian_delta_max} show that regardless of the amplitude and the modulation depth $\Delta_{max}$ the super-Gaussian pulse offers a higher degree of robustness.
	
\begin{figure}
		\centering
		\includegraphics[width=0.5\textwidth]{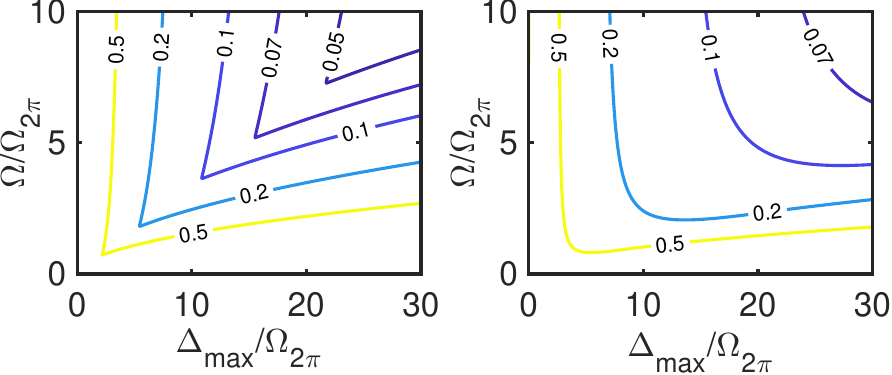}
		\caption{The maximum value of $\eta$ for a pulse with a super-Gaussian (left) and a Gaussian envelope (right).
		}
		\label{fig:adiabaticity_gaussian_vs_supergaussian}
	\end{figure}

	Additionally, one can study the adiabaticity of the pulses according to the criterion $\eta = \frac{\| \dot{\Omega}\Delta - \Omega\dot{\Delta}\|}{ (\Omega^2 + \Delta^2)^{3/2}}$, for an adiabatic trajectory $\eta\ll1$ \cite{Vitanov2001}.
	Fig.~\ref{fig:adiabaticity_gaussian_vs_supergaussian} shows the maximum value $\rm{max}_t(\eta)$  as a function of $(\Omega,\Delta_{\rm max})$ for the Gaussian and super-Gaussian envelope.
	The super-Gaussian pulse has a similar average value of $\eta$ (not shown), but a lower peak value, which implies an adiabatic trajectory, as was claimed in the main text.
	
	\section{Arbitrary $X/Y$ Bloch sphere rotations}

	\subsection{Arbitrary rotations based on the modulated pulses} \label{app:arb_rot_int}
	
	Without losses, the time evolution operator $U$ fully describes the effect of a gate on a qubit. In order to analyze the effect of the modulated drive, here it is more convenient to work in the frame co-rotating with the qubit (and not the pulse, as presented in the main text). Then the Hamiltonian is given by $H = \ket{e}\bra{g} \Omega e^{i\phi(t)} + h.c.$, where $\phi(t) = (\omega_d(t) - \omega_q)t$. With the initial condition $U(-\frac{T}{2})=I$ the equation of motion for $\dot{U}= -i H U$ can be integrated to obtain $U(\frac{T}{2})$.		
	
	Any $U$ can be decomposed into linear combinations of $\sigma_0=I$, $\sigma_x$, $\sigma_y$ and  $\sigma_z$: $U = c_0 \sigma_0 - i \sum_{j=1}^3 c_j \sigma_j$.
	For a rotation by angle $\theta$ generated by $\sigma_x$ the corresponding evolution operator is given by $U = \begin{pmatrix}
	\cos(\theta/2) & i \sin(\theta/2) \\ i\sin(\theta/2) & \cos(\theta/2)	\end{pmatrix} = 	\cos(\theta/2)\sigma_0 + i \sin(\theta/2)\sigma_x$. If the plane of rotation is off, i.e. the generator of rotation is a linear combination of $\sigma_x$ and	$\sigma_y$, there will be a nonzero $\sigma_y$ component in $U$.		
	
	The goal of the second part of this work was to generate a frequency modulated pulse which is effectively a rotation generated by $\sigma_x$, while simultaneously being insensitive to amplitude deviations. 
	
\begin{figure}[h!]
		\centering
		\includegraphics[width=\defaultfigurewidth]{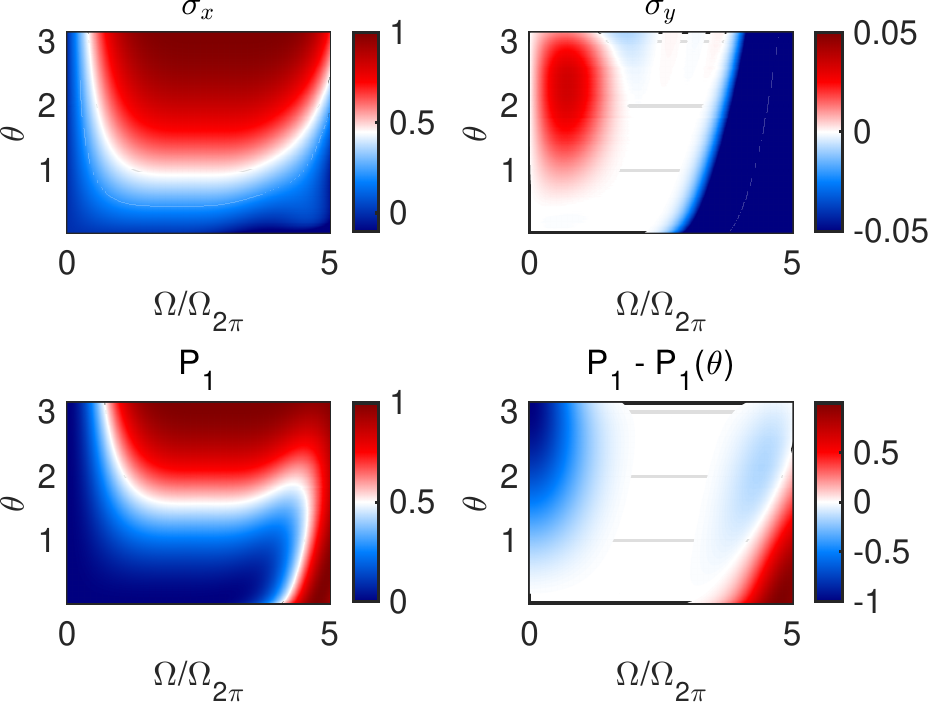}
		\caption{The $\sigma_x$ (top left) and $\sigma_y$ (top-right) components of $U$, the population of the excited state (bottom left) and the deviation from the desired value $P_1(\theta) = \sin(\theta/2)^2$ (bottom right) as a function of $\theta$ and $\Omega$.}
		\label{fig:arb_pulses_interpolation_decompose}
	\end{figure}

	In the main text this was demonstrated using quantum process tomography. Here we present a diagram which completely characterizes our pulses as a function of the rotation angle $\theta$ and their amplitude $\Omega$. The evolution operator is calculated; Fig.~\ref{fig:arb_pulses_interpolation_decompose} shows the $\sigma_x$ and $\sigma_y$ components of $U$, along with the population of the excited state $P_1$, and its deviation from the desired value $P_1(\theta) = \sin(\theta/2)^2$. We see that in the amplitude-robust region (i.e. where $P_1 \approx \sin(\theta/2)^2$) the $\sigma_y$ component is negligible and the $\sigma_x$ component is what generates the rotation. Additionally, this map was computed by interpolating pulse parameters between the 10 optimized ones (for $P_i = i/10$, where $i\in{1,...,10}$), and shows that an amplitude robust arbitrary rotation can be obtained this way.

	\subsection{Detuning robustness of arbitrary rotations}\label{app:trig_det_rob}
	
\begin{figure}[h!]
		\centering
		\includegraphics[width=\defaultfigurewidth]{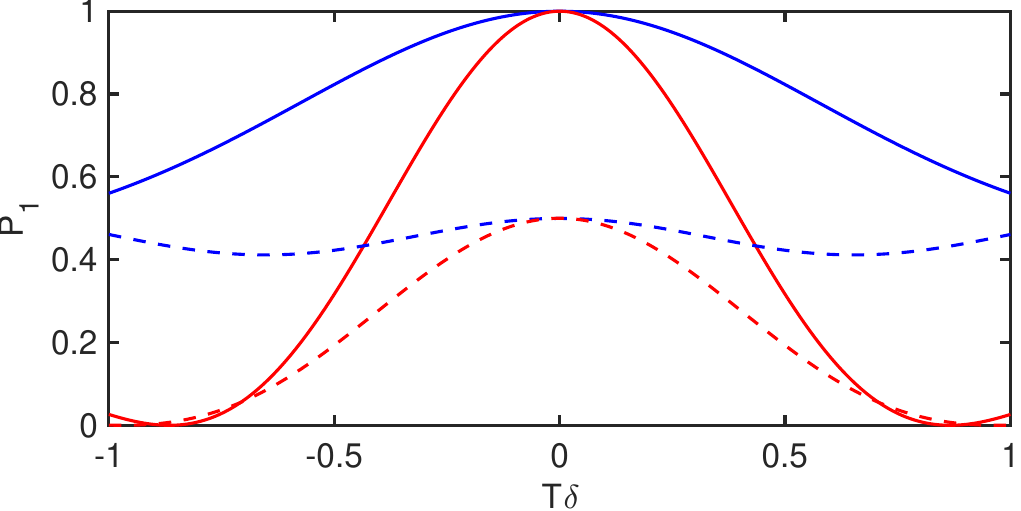}
		\caption{The population of the excited state after applying the robust (blue) and Rabi (red) $\pi/2$ (dashed line) and $\pi$ (solid line) pulses, to a qubit initialized in the ground state, as a function of detuning $\delta$.}
		\label{fig:detuning_robustness_arb}
	\end{figure}

	Here we study the effect of finite detuning, i.e. an error in the pulse frequency, of the pulses presented in section \ref{rotation} of the main text.
	We present the results in the same manner as Fig.~\ref{fig:supergaussian_amp_and_freq} of the main text as well as the supplementary section \ref{app:sg_vs_rect_env}. Fig.~\ref{fig:detuning_robustness_arb} compares the modulated and the usual Rabi pulses for target angles $\theta=\pi$ and $\theta=\pi/2$: although both deviate from the desired value at non-zero detunings, the Rabi ones are more affected by it. It is worth reiterating that detuning robustness was not a design goal for these pulses, as for $\delta\neq0$ the qubit phase evolves as $\dot{\phi} = \delta$, and the (vertical) plane of rotation is not easily controllable. Nevertheless, the results presented here show that the modulated pulses are more resilient to frequency errors than the usual Rabi ones.

	\subsection{Comparison with composite pulses}
	
\begin{figure}[h!]
		\centering
		\includegraphics[width=\defaultfigurewidth]{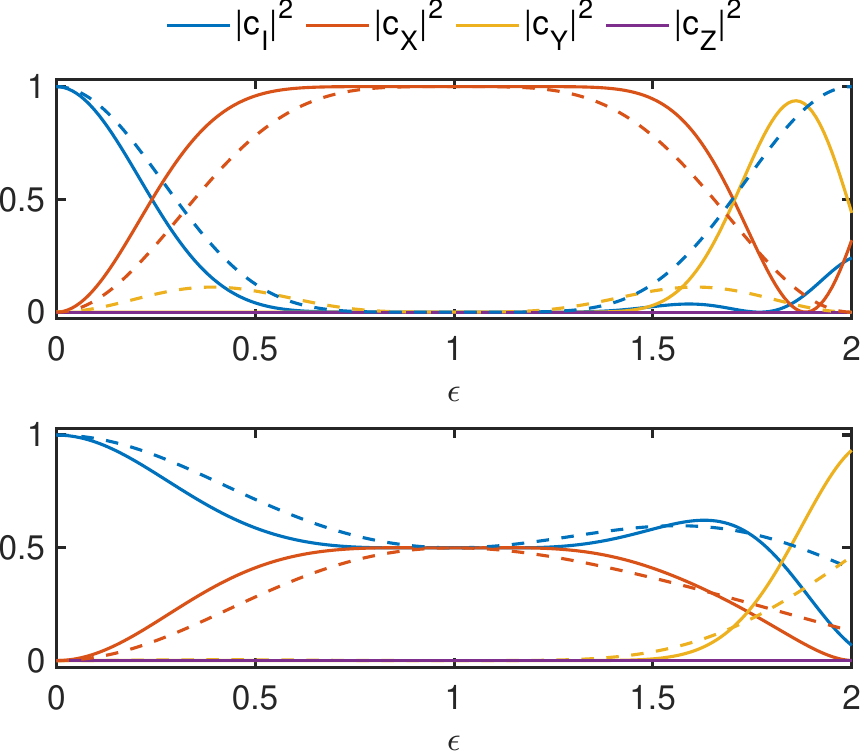}
		\caption{The expansion of the propagator $U= \sum_{i} \sigma_i c_i$, $i \in\{I,X,Y,Z\}$ as a function of the relative amplitude $\epsilon = \Omega/\Omega_0$ for our modulated pulse (solid lines) and a 3-pulse composite (dashed). Two $\theta$'s are shown: $\pi$ on the top panel and $\pi/2$ on the bottom one.}
		\label{fig:trig_vs_CP}
	\end{figure}

	Here we perform a comparison of our pulses to a short composite pulse presented in \cite{kukita2022short}, engineered to provide amplitude robustness.
	As the sequence consists of only 3 consecutive pulses, the total duration of the composite pulse should be comparable to the pulses presented in this work.
	Fig.~\ref{fig:trig_vs_CP} shows the expansion of the propagator $U$, in terms of the Pauli matrices, as a function of the relative amplitude $\epsilon = \Omega/\Omega_0$ ($\Omega_0$ is the optimal amplitude). Our pulses outperform the composite one, the working range is wider and flatter. This is especially evident in the case of a $\pi/2$ pulse, where the composite pulse deviates approximately quadratically $(\epsilon-1)^2$ from the target operation, while the modulated one is plateau-like around $\epsilon=1$. Longer pulse sequences might offer more robustness, at the expense of a longer duration. Finding such sequences is also not trivial: in the same work \cite{kukita2022short} it was also shown that the 3-pulse sequence outperforms a 6-pulse one.
	
	\FloatBarrier

\end{document}